%% file: main.tex
\documentclass[10pt, aps,prc,twocolumn,superscriptaddress,preprintnumbers,
amsmath, 
floatfix,
longbibliography,
nofootinbib
]{revtex4-1}
\usepackage[T1]{fontenc}
\usepackage[utf8x]{inputenc} 
\usepackage{adjustbox}          
\usepackage[caption=false]{subfig}
\usepackage{url}
\usepackage{color}
\usepackage{float}
\usepackage[pdftex,colorlinks=true, linkcolor = blue, citecolor=blue,urlcolor=blue, bookmarksnumbered=true, bookmarksopen=true]{hyperref}
\usepackage{longtable}
\usepackage{amsfonts}
\usepackage{dsfont}
\usepackage{wrapfig,bm} 
\usepackage[normalem]{ulem}
\usepackage{MnSymbol}

\begin{document}
\title {Non-Equilibrium Dynamics of Hard Spheres in the Fluid, Crystalline, and Glassy Regimes}
\author{Matthew Kafker}
\affiliation{Department of Physics, University of Washington, WA, USA}
\author{Xerxes D. Arsiwalla }
\affiliation{Wolfram Institute for Computational Foundations of Science, IL, USA}

\date{\today}

{\let\thefootnote\relax 
  \footnotetext{Corresponding Author: M. Kafker, Email: \href{mailto:kafkem@uw.edu}{kafkem@uw.edu}  \\
  Author X. D. Arsiwalla, Email:  \href{mailto:x.d.arsiwalla@gmail.com}{x.d.arsiwalla@gmail.com} }
}

\preprint{NT@UW-25-6}
\begin{abstract}

We investigate the response of a system of hard spheres to two classes of perturbations over a range of densities spanning the fluid, crystalline, and glassy regimes within a molecular dynamics framework. Firstly, we consider the relaxation of a ``thermal inhomogeneity,'' in which a central region of particles is given a higher temperature than its surroundings and is then allowed to evolve under Newtonian dynamics. In this case, the hot central ``core'' of particles expands and collides with the cold surrounding material, creating a transient radially-expanding ``compression wave,'' which is rapidly dissipated by particle-particle collisions and interactions with periodic images at the boundary, leading to a rapid relaxation to equilibrium. Secondly, we consider a rapid compression of the spheres into a disordered glassy state at high densities. Such rapidly compressed systems exhibit very slow structural relaxation times, many orders of magnitude longer than thermalization times for simple temperature inhomogeneities. We find that thermal relaxation of the velocity distribution is determined simply by the total collision rate, whereas structural relaxation requires coordinated collective motion, which is strongly suppressed at high density, although some particle rearrangement nevertheless occurs. We further find that collisions propagate significantly faster through glassy systems than through crystalline systems at the same density, which leads to very rapid relaxation of velocity perturbations, although structural relaxation remains very slow. These results extend the validity of previous observations that glassy systems exhibit a hybrid character, sharing features with both equilibrium and non-equilibrium systems. Finally, we introduce the hard sphere causal graph, a network-based characterization of the dynamical history of a hard sphere system, which encapsulates several useful metrics for characterizing hard sphere systems within a single structure, and which emphasizes the role of causality in these systems.


\end{abstract} 

\maketitle

\section{Introduction}
Hard spheres (HS) are a widely used model system for gases, liquids, solids, colloids, glasses, and granular materials. They are perhaps the longest studied interacting many-body system in the physics literature \cite{mulero2008theory}, with published results dating back over 150 years, at least to the Ph.D. dissertation of van der Waals in the 1870's. Although the collision dynamics of hard spheres are simple, in the sense that the result of a two particle collision under Newton's laws can be computed exactly, HS systems containing many particles nevertheless exhibit a variety of interesting and surprising behaviors, both static and dynamic, which are difficult to explain analytically. Much has been understood about HS systems, across the entire phase diagram, since the pioneering work of Alder and Wainwright in the 1950's \cite{alder1957phase,alder1959studies1,alder1960studies}, but surprisingly, more is still being discovered to this day, in part due to the intense numerical calculations required to investigate HS systems in certain regimes, which were therefore inaccessible in these early studies, but also due to several theoretical advances. Hard spheres are of interest to the modern condensed matter physics literature primarily due to their prominent role in studies of the glass transition \cite{parisi2010mean,charbonneau2017glass,parisi2020theory,berthier2011theoretical}, a problem which P. W. Anderson described as ``the deepest and most interesting problem in solid state theory'' \cite{anderson1995through}, and their related role in the study of the jamming transition \cite{van2009jamming} and the Gardner transition \cite{berthier2019gardner}. It is also possible now to perform experiments with colloidal systems which have been carefully tuned so as to accurately approximate hard sphere dynamics \cite{royall2024colloidal}, and which will therefore offer a valuable synergistic perspective on this crucial model system.

Apart from very low density configurations which are nearly ideal-gas like, and very high density configurations which are nearly perfectly regular crystals, the behavior of HS systems is difficult to calculate analytically, and primarily one relies on numerical methods to characterize the systems in this intermediate regime (which spans much of the phase diagram). At low densities, the equation of state (EOS) may be approximated by the virial expansion \cite{mulero2008theory} or fit with an approximate analytical formula such as the Carnahan-Starling EOS \cite{carnahan1969equation}, both of which offer an accurate description of the pressure up to a certain density; however, neither predicts the existence of the fluid-solid transition, and the Carnahan-Starling EOS predicts a maximum packing fraction of $\phi=1$, which exceeds the maximum close packing density of monodisperse spheres in three dimensions, $\phi_\text{FCC}\approx0.74$, associated with the face-centered cubic (FCC) lattice \cite{conway2013sphere}, and where $\phi=\frac{4\pi R^3 }{3}\frac{N}{V}$ is the packing fraction of $N$ spheres with radius $R$ in a volume $V$. Numerical methods are therefore preferable, and they split largely into two classes: Monte Carlo methods \cite{metropolis1953equation,hansen2013theory,bernard2009event,engel2013hard} and molecular dynamics methods \cite{hansen2013theory,alder1959studies1,smallenburg2022efficient}. Monte Carlo methods perform random trial movements of spheres and accept these moves provided no overlaps occur; these methods do not propagate particles along Newtonian trajectories and are therefore restricted to investigations of equilibrium physics. Molecular dynamics methods, by contrast, propagate particles along Newtonian trajectories, and are therefore useful for investigating both equilibrium and non-equilibrium phenomena. 

Despite the reliance on numerics, the equilibrium physics of hard spheres in three dimensions, as expressed in the equation of state, is relatively well characterized. Due to the singular nature of the HS interaction potential, the temperature has a trivial effect on the spatial configurations of the spheres, which determine the (reduced) pressure \cite{mulero2008theory}, and so the equation of state depends on a single parameter, the packing fraction $\phi$. We briefly describe here the qualitative structure of the hard sphere phase diagram in three dimensions, see for example Fig. 3 of \textcite{parisi2010mean}. Starting from a low density, the system is in a fluid state. Upon increasing the density (infinitesimally slowly, so that the system remains in equilibrium), a first-order phase transition occurs (in the thermodynamic limit), and the system freezes into a crystalline state. As the density increases further, the system will remain a crystal until the pressure diverges at the FCC density. 

Non-equilibrium physics of hard spheres, the focus of this paper, covers much wider territory and is correspondingly less well understood. The term ``non-equilibrium'' covers many cases, and we shall focus on just two in this study. However, the analysis techniques we employ below were deliberately chosen because they are very general and they can be straightforwardly applied to a wide variety of non-equilibrium processes in HS systems; we will return to the question of more general non-equilibrium dynamics in the conclusion of the paper. The two cases we consider in detail in this study are as follows. The equilibrium velocity distribution for HS is the Maxwell-Boltzmann distribution, well known from elementary statistical physics. One generic variety of non-equilibrium behavior is obtained when the velocity distribution is forced to take on a different character. If the spheres are prepared in this manner, the system generally returns to equilibrium, but how does the relaxation occur, in detail? Another simple ``perturbation'' away from equilibrium can be accomplished by compressing the system at a finite rate, or equivalently by growing the spheres in a fixed simulation volume, until a desired density is achieved. If one prepares the system in this manner, at low post-compression densities, the physics (i.e., the pressure) is unchanged; however, upon reaching a critical density (very close to the onset of solidification \cite{skoge2006packing}), a bifurcation of the equation of state occurs, in which systems which are compressed sufficiently rapidly are trapped in a thermodynamically metastable, spatially disordered state. These states exhibit a higher pressure when compared to a configuration which is compressed infinitesimally slowly and is thus crystalline in character, at the same density and energy. The states on the metastable branch are ``glassy'' in character, and despite their metastability, they do not generally relax to equilibrium in a finite duration simulation. Why do these configurations not relax to equilibrium under molecular dynamics evolution, while the ``thermal perturbations'' do? What do these systems do if they are allowed to evolve in time, if they do not thermalize? 

In this study, we investigate these two varieties of perturbation and try to address these questions. 
In Section \ref{sec:EDMD}, we outline the event-driven molecular dynamics (EDMD) simulation algorithm, the only approach capable of simulating the entire HS phase diagram, both in and out of equilibrium. In Section \ref{sec:ThermRel}, we investigate perturbations of the velocity distribution away from the ground state branch of the equation of state and their relaxation to equilibrium. In Section \ref{sec:StructRel}, we investigate the effects of rapid compression on the system of HS and the physics on the metastable glassy branch of the phase diagram. Sections \ref{sec:ThermRel} and \ref{sec:StructRel} constitute the main scientific results of the paper. In the following two sections (\ref{sec:CG} and \ref{sec:CGForGlassyStates}), we present an alternative perspective on the results presented in Sections \ref{sec:ThermRel} and \ref{sec:StructRel} by introducing a network-based characterization of HS systems that emphasizes the role of causality in these systems.  In Section \ref{sec:CG}, we introduce the hard sphere causal graph, a directed acyclic graph derived from the dynamical evolution of the HS system, and which we argue can be used to characterize certain aspects of the equilibrium and non-equilibrium physics of the HS systems. In Section \ref{sec:CGForGlassyStates}, we use the hard sphere causal graph to explore more deeply the relationship between the spatial disorder of glassy states and their enhanced pressure relative to crystalline states at the same density, thereby extending the results presented in Section \ref{sec:StructRel}.  Finally, in Section \ref{sec:conclusion}, we summarize our findings.

\section{Event-Driven Molecular Dynamics}
\label{sec:EDMD}
Our study requires the use of two distinct but related codes, corresponding to preparing initial conditions and evolving the system in time, respectively. We offer a brief summary of both simulation methods here. 

For dynamics, we follow  \textcite{smallenburg2022efficient} (and the references therein) and implement an efficient event-driven molecular dynamics (EDMD) code, which propagates the spheres along Newtonian trajectories in time with high accuracy, and which leverages a cell list and an event calendar for rapid determination of updates to the system \cite{githubrepo}. For studies of crystallization, glassy dynamics, and more generally non-equilibrium dynamics at high density, having a highly efficient algorithm is essential, as it is not uncommon to perform simulations with millions or even billions of time steps. We therefore dedicate a discussion to our implementation details, although it closely follows established procedures in the literature. (Experienced practitioners may therefore safely skim this section, or even skip it altogether.) We note that our code is highly accurate, with momentum and kinetic energy being conserved to machine precision over hundreds of millions of steps.

In EDMD, the particles are moved along Newtonian trajectories. For the hard sphere interaction, this means the particles move along linear paths with constant velocities until a collision occurs. The collision dynamics can be implemented using the following simple rule, which follows from kinetic energy and momentum conservation. 

\begin{equation}\label{Eq:VelUpd}
\begin{aligned}
    \textbf{v}_1'=\textbf{v}_1^\perp+\textbf{v}_2^\parallel, \\
    \textbf{v}_2'=\textbf{v}_2^\perp+\textbf{v}_1^\parallel,
    \end{aligned}
\end{equation}
where the primed variables refer to the velocities after the collision. Stated simply, at the moment of collision, the velocities are decomposed into components perpendicular and parallel to the axis connecting the centers of the two spheres. Along this axis, the components are exchanged during the collision. Perpendicular to this axis, the velocity components are unchanged. To determine the time of the next collision, we observe that in molecular dynamics, the spheres move in straight lines when they are not colliding. So for two spheres of radius $R$ currently with states $(\textbf{r}_1,\textbf{v}_1)$ and $(\textbf{r}_2,\textbf{v}_2)$, a collision will take place when the following holds:
\begin{gather}
    \vert \textbf{r}_1+\textbf{v}_1 \Delta t - \textbf{r}_2-\textbf{v}_2 \Delta t\vert^2=4R^2, 
\end{gather}
which is satisfied when 

\begin{gather}\label{Eq:Time}
    \Delta t=\frac{1}{v_{12}^2}\bigg(-\textbf{r}_{12}\cdot\textbf{v}_{12}-\sqrt{(\textbf{r}_{12}\cdot\textbf{v}_{12})^2- v_{12}^2(r_{12}^2-4R^2)}\bigg),
\end{gather}
where $\textbf{r}_{12}\equiv\textbf{r}_1-\textbf{r}_2$ and $\textbf{v}_{12}\equiv\textbf{v}_1-\textbf{v}_2$. In solving the quadratic equation for $\Delta t$, we have taken the negative root to select the shorter of the two possible collision times, which will correspond to the point when the particles are separated by distance $2R$ and where the lines $\textbf{r}_1(t)$ and $\textbf{r}_2(t)$ are converging in 3D (i.e., rather than diverging). If two particles will not collide, then $\textbf{r}_{12}\cdot\textbf{v}_{12}\geq0$ or the argument of the square root will be negative. 

The EDMD simulation can be implemented by using Eq. \eqref{Eq:Time} to determine the next collision time for all particles (which will be accurate to machine precision), then evolving them all along their velocity vectors by the shortest such time, $\Delta t$. The particles which collide have their velocities updated according to Eq. \eqref{Eq:VelUpd}. However, a na\"{i}ve implementation of this procedure can be highly inefficient, for example requiring $N(N-1)/2$ particle-particle distance checks per time step, as well as enforcing periodic boundary conditions. We next describe the steps we have taken to improve efficiency.

Firstly, we have implement a ``cell list'' \cite{smallenburg2022efficient}, which refers to a domain decomposition of the simulation box into small cubic cells. At all times, we store the cell in which each particle currently resides. If the cell size is chosen appropriately (i.e., to be a sphere diameter in sidelength), we only need to consider possible collisions with particles in the 27 neighboring cells, as well as the times that each particle will cross a cell boundary. In this way, the set of possible next ``events'' is greatly reduced.

As the set of possible events is constructed, the events are loaded into the ``event calendar,'' a data structure which is specifically chosen to automatically and efficiently sort them by time (e.g., a binary search tree \cite{smallenburg2022efficient}), a feature which is particularly useful as the set of possible future events can be very large in EDMD calculations. Addition or deletion of individual elements using these automatic-sorting data structures scales as $\mathcal{O}(\log(N_\text{elements}))$, see for example the \verb|SortedSet| datatype in the \verb|sortedcontainers| Python library \cite{sortedcontainers}. The next event, in the sense of the smallest $\Delta t$, is then ``popped'' from the front of the event calendar and executed. If this event is a cell crossing, then we update the particle's position and cell information. If it is a collision, then the positions and velocities of the colliding particles are updated. Greater efficiency still can be obtained by only updating the event calendar for particles which were involved in the most recent event. That is, the future events are only recomputed for the one particle involved in the cell crossing or the two particles involved in the collision. All other future events for other particles are unchanged, because only the shortest-time event has taken place, and any other events would necessarily occur later in time. In this way, we make fewer insertions or deletions into the event calendar, and we avoid redundantly recomputing future events. To accommodate this optimization, we have each particle store the time at which it last moved, so that each particle's state is only ever updated when an event involving that particle is popped from the event calendar. It is always possible to synchronize the global state of the system by moving each particle along its velocity vector until that particle's individual time variable matches the latest time of the system. This can be done without risk of cell crossings or collisions, because if such events were to occur before the latest update, they would have already been accounted for in the event calendar and executed.

We next turn our attention to the preparation of initial conditions. At low densities, the task of preparing an initial state is simple. One can simply place particles in random locations, or in a regular grid, and provided no overlaps have taken place, there is nothing more to be done. However, at high densities, more care must be taken. Of course, provided that the density is not too high, one can always start spheres in face-centered cubic or hexagonal close-packed configurations, which are known to be the densest packing of spheres in 3D \cite{conway2013sphere}. However, for the study of glassy dynamics, where disordered high-density states are desired, this is insufficient, as at high enough densities a system will effectively never evolve from a  crystalline configuration to a disordered configuration, or vice versa, and randomly placing spheres to obtain high-density amorphous configurations is impractical (in fact, it is impossible at high enough densities \cite{skoge2006packing}). We therefore desire a framework through which high density crystalline and amorphous configurations can both be obtained on equal footing, ideally by changing a single parameter.

One framework which meets our criteria is the Lubachevsky-Stillinger (LS) procedure \cite{lubachevsky1990geometric}, which is closely related to the EDMD procedure described above, and can thus use all of the optimizations we implemented previously. The LS procedure is a protocol in which spheres are started as point particles randomly distributed in the box with random initial velocities. The diameter of the spheres is then grown during the dynamical evolution of the system, until some specified density is achieved or the spheres jam up and the collision rate diverges.

For LS calculations, trajectories are Newtonian, just as in EDMD, so between collisions, particles move along straight lines. However, the particles are now growing in volume, in that the sphere diameters $D$ grow linearly with time as 

\begin{equation}\label{Eq:Diam}
    D(t) = 2R(t) =\gamma t.
\end{equation}  The growing surface of each particle now carries with it kinetic energy, and hence the kinetic energy is no longer conserved during the collision, and the new velocity update rule is given by

\begin{equation}\label{Eq:VelUpdLS}
\begin{aligned}
    \textbf{v}_1'=\textbf{v}_1^\perp+\textbf{v}_2^\parallel + \gamma \hat{\textbf{r}}_{12}, \\
    \textbf{v}_2'=\textbf{v}_2^\perp+\textbf{v}_1^\parallel - \gamma \hat{\textbf{r}}_{12},
    \end{aligned}
\end{equation}
where $\hat{\textbf{r}}_{12}=\textbf{r}_{12}/r_{12}$ is a unit vector.
Finally, an updated collision time formula must be derived using

\begin{gather}
    \vert \textbf{r}_1+\textbf{v}_1 \Delta t - \textbf{r}_2-\textbf{v}_2 \Delta t\vert^2=4R(t)^2. 
\end{gather}

The parameter $\gamma$ can be used to prepare the system in different states. The spheres can be made to crystallize by using very low values of $\gamma$, $\gamma\approx10^{-5}$ or less \cite{skoge2006packing}. Anything above this, the system will not have sufficient time to crystallize, and will be trapped in a disordered configuration.

\section{Relaxation of Thermal Inhomogeneities}
\label{sec:ThermRel}

To characterize the non-equilibrium dynamics of hard spheres in relatively simple circumstances, we consider a situation similar in spirit to the impulse-response behavior of the heat equation, in which a localized region is initially given a higher temperature than the surroundings, and the system is then allowed to thermalize. We prepare a system of $N=4096$ hard spheres at various densities $\phi \in [0.08, 0.62]$ with periodic boundary conditions, choose a central spherical region containing approximately $5\%$ of particles, and give them a higher ``temperature'' by increasing their speed by a factor of 10 relative to the other particles. (All simulations are performed at the same total energy.) We then evolve this system using the EDMD procedure described above, and the velocity distribution relaxes to the equilibrium distribution, which is a Maxwell-Boltzmann distribution. The initial conditions for all calculations in this section were prepared using the LS procedure with a slow compression rate of $\gamma = 10^{-5}$ which allows the high-density systems sufficient time to crystallize \cite{skoge2006packing}.  

\begin{figure}
    \centering
    \includegraphics[width=1.0\columnwidth]{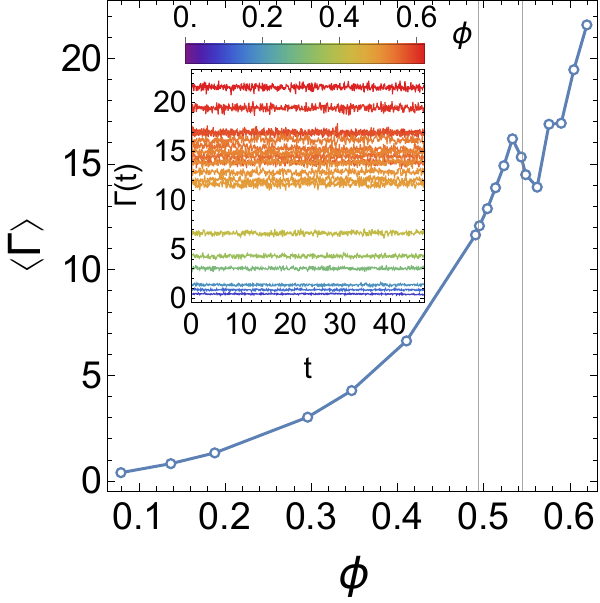}
    \caption{The average single-particle collision rate $\langle\Gamma\rangle$, which is related to the pressure via an affine transformation \cite{alder1960studies,erpenbeck1984molecular,hoover1967studies,engel2013hard}, as a function of the packing fraction, $\phi = \frac{4\pi R^3}{3}\frac{N}{V}.$ The vertical gray lines indicate the density of freezing, $\phi_\text{freeze}\approx 0.494$, and the density of melting, $\phi_\text{melt}\approx 0.545$, as reported in, for example, \cite{skoge2006packing}. Between these densities, the fluid and solid phases can coexist in the thermodynamic limit, although one or the other typically dominates for finite system sizes \cite{alder1957phase,skoge2006packing}. Our system appears to fully crystallize around $\phi \approx 0.562$. In the inset, we see the time evolution of the average single particle collision rate, which exhibits a stable average at all densities, with small fluctuations, indicating that the system is at equilibrium. Note that in the main figure, $\langle \Gamma \rangle$ is averaged over all particles and over all times, whereas in the inset, $\Gamma(t)$ is averaged over all the particles.}
    \label{fig:SPCRInset}
\end{figure}

For reference, we also prepare a system at each density where the velocities are prepared in the equilibrium distribution (but at the same total energy as the non-equilibrium experiments). We show the results for the single-particle collision rate, averaged over all particles and over time, in Fig. \ref{fig:SPCRInset}. The average single-particle collision rate is related to the pressure, and hence the equation of state, by an affine transformation \cite{alder1960studies,erpenbeck1984molecular,hoover1967studies,engel2013hard}, and is therefore a fundamental characterization of the equilibrium properties of the hard sphere system \cite{hansen2013theory}. (Note that these authors use $\Gamma$ to refer to the \textit{total} collision rate, but here we take $\Gamma$ to mean the \textit{single-particle} collision rate, as we will emphasize aspects of single-particle motion in this study. These two quantities are related by a trivial rescaling.) As was first noted by Alder and Wainwright \cite{alder1957phase} in the 1950's, starting from a low-density configuration, the pressure is increased until a first-order phase transition from a fluid to a solid phase is observed, which we see here from the presence of a non-monotonicity in the pressure vs. density curve. A drop in the pressure is observed, which is consistent with our expectations for crystallization, where the spatial organization of spheres takes on a regular crystalline character, and the instantaneous collision rate is lower than for a spatially disordered system at the same density \cite{parisi2010mean}. Our system appears to fully crystallize at $\phi\approx0.562$. 

We note that the precise determination of the HS equation of state is not our goal here, as this has been done by numerous authors elsewhere, see, for example, \textcite{mulero2008theory} (and references therein) for low densities and \textcite{skoge2006packing,speedy1998pressure,speedy1997pressure} for high densities. Instead, we are simply demonstrating that the single-particle collision rate can be used to study the equilibrium physics of hard sphere systems. Furthermore, we shall use Fig. \ref{fig:SPCRInset} as a reference behavior for the analysis of perturbations, as the collision rate is also a simple measure of the non-equilibrium dynamics of the system. 

\begin{figure}
    \centering
    \includegraphics[width=1.0\columnwidth]{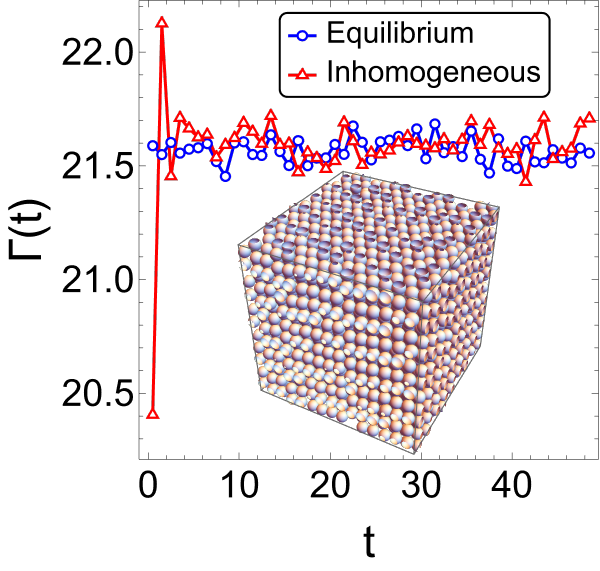}
    \caption{Average single particle collision rate vs. time for a system in which a central spherical region containing about 5\% of the spheres is initially given a higher temperature than the surrounding region (``inhomogeneous,'' red points). More precisely, the spheres in this central ``hot'' region are given 10 times the initial speed of the particles in the surrounding ``cold'' region. For comparison, a system is prepared with the same initial positions and the same energy, but all particles are given speeds consistent with a Maxwell-Boltzmann distribution (``equilibrium,''  blue points). The spike in the collision rate at early times arises from the hot central core of particles slamming into the cold surroundings, creating a short-lived longitudinal ``compression wave,'' which is quickly smoothed out by particle-particle collisions and the periodic boundaries. The initial positions are prepared at density $\phi=0.62$ in a crystalline state. The inset shows the configuration of spheres in space with periodic boundary conditions, prepared using the LS procedure with $\gamma=10^{-5}$.}
    \label{fig:FastRel}
\end{figure}

\begin{figure}
    \centering
    \includegraphics[width=1.0\columnwidth]{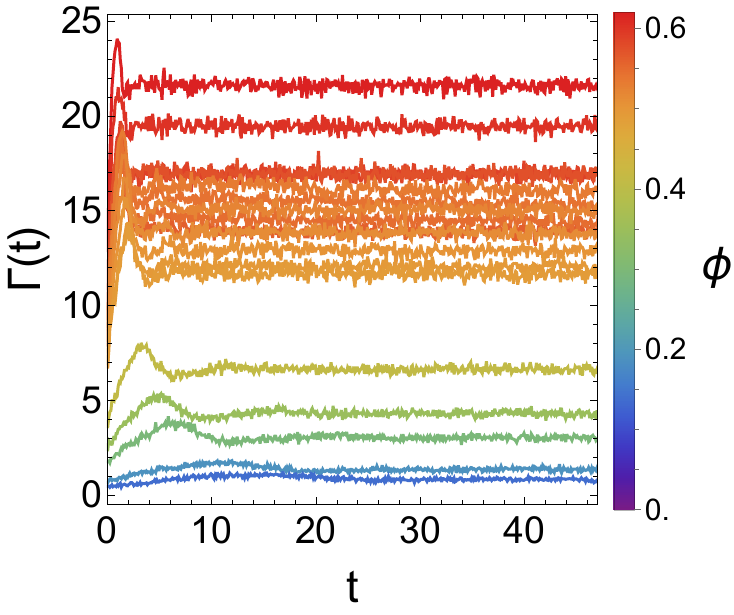}
    \caption{The same setup as Fig. \ref{fig:FastRel}, except only the ``inhomogeneous'' curves are shown. Here, we repeat this experiment with initial conditions spanning a range of densities, covering  low-density fluid to high-density crystalline configurations, see colorbar. The system relaxes to  equilibrium after some transient behavior. The transient exhibits some density dependence.}
    \label{fig:SPCRThermRel}
\end{figure}

We are now prepared to consider the results of the velocity perturbations described above. When the HS system is heated in the center at a particular density, the particles in the hot central region slam into the cold surroundings, redistributing the kinetic energy until equipartition is restored, and the velocity distribution is one again Maxwell-Boltzmann in character. These relaxation dynamics can be seen in the behavior of the average single-particle collision rate, see Fig. \ref{fig:FastRel}. An initial transient is observed in the collision rate, which quickly relaxes to fluctuate around the same equilibrium value that the uniform system does. The evolution of this relaxation behavior with density can be seen in Fig. \ref{fig:SPCRThermRel}. At all densities, this transient can be observed, and its duration and profile depend on the density.

\begin{figure}
    \centering
    \includegraphics[width=1.0\columnwidth]{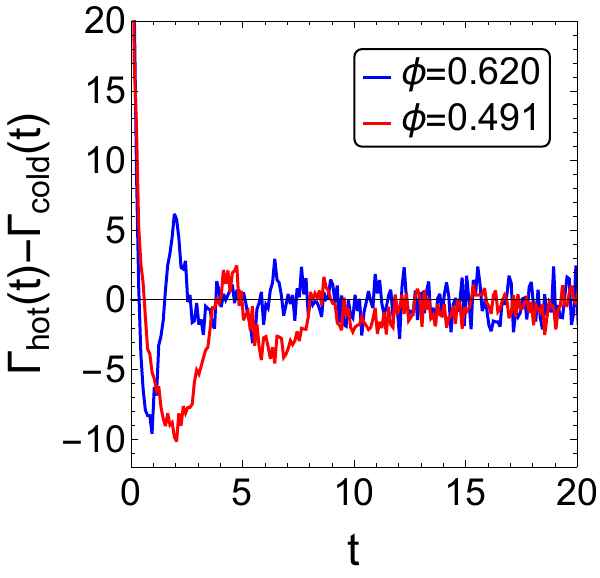}
    \caption{The difference in average single-particle collision rates between the initially ``hot'' and initially ``cold'' subsets of particles, see Figs. \ref{fig:FastRel} and \ref{fig:SPCRThermRel}. The red curve shows a system just below freezing density in the fluid regime, and the blue curve is well into the crystalline solid regime. Positive values indicate the subset of particles initially given higher ``temperature'' has a higher instantaneous single-particle collision rate than the subset of initially ``cold'' particles, and negative values indicate the opposite situation. The oscillation of this curve at early times constitutes evidence for a radially expanding longitudinal ``compression wave.'' In the crystalline regime, particles are packed more tightly, and hence the ``compression wave'' propagates faster through the system, producing more particle-particle collisions per unit time, and hence the system relaxes to equilibrium faster.}
    \label{fig:HotMinusCold}
\end{figure}

We can understand this thermal relaxation behavior in more detail. In particular, we can isolate the initially ``hot'' particles and, separately, isolate the initially ``cold'' particles (i.e., initially high speed and low speed particles, respectively), and then we can then separately calculate the average single-particle collision rate within those groups over time, which we shall call $\Gamma_\text{hot}(t)$ and $\Gamma_\text{cold}(t)$ respectively. The difference $\Gamma_\text{hot}(t)-\Gamma_\text{cold}(t)$ offers a more detailed picture of the relaxation dynamics, see Fig. \ref{fig:HotMinusCold}. While the initially ``hot'' particles start with a much higher collision rate (as expected), and so $\Gamma_\text{hot}(t)-\Gamma_\text{cold}(t)>0$ at early times, that kinetic energy is rapidly imparted to the rest of the system, so $\Gamma_\text{hot}(t)-\Gamma_\text{cold}(t)\rightarrow0$ eventually. The oscillation of $\Gamma_\text{hot}(t)-\Gamma_\text{cold}(t)$ at early times constitutes evidence for a radially expanding ``compression wave'' in the material, as the initially hot particles slam into the cold surroundings, dumping their kinetic energy into the surrounding material and exerting a pressure radially outward on the spherical ``shell'' of material surrounding the hot core. This pressure compresses the spheres in the surrounding spherical shell, locally increasing the density and the collision rate, and the spheres in that shell in turn expand into the surrounding material in an approximately spherical front. Because of the HS collision dynamics, see Eq. \eqref{Eq:VelUpd}, and the fixed total particle number, this radially expanding compressed region temporarily leaves a lower density, lower collision rate, region behind it. We note that the presence of this ``compression wave'' in the dynamics indicates an important difference between the impulse-response behavior of the heat equation and the thermal relaxation of the hard sphere system.

\begin{figure}
    \centering
    \includegraphics[width=1.0\columnwidth]{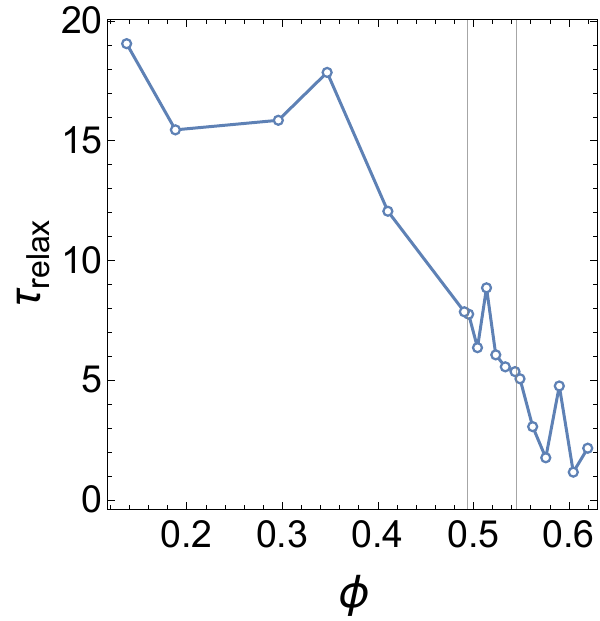}
    \caption{The relaxation time of temperature inhomogeneities as a function of the density for the setup in Figs. \ref{fig:FastRel}, \ref{fig:SPCRThermRel}, and \ref{fig:HotMinusCold}. The relaxation time is defined to be the point when $\Gamma_\text{hot}(t)-\Gamma_\text{cold}(t)$ is statistically consistent with zero, see Fig. \ref{fig:HotMinusCold}, at which point the velocity distribution is again a Maxwell-Boltzmann distribution. Vertical lines indicate the fluid-solid coexistence region in the thermodynamic limit, see caption of Fig. \ref{fig:SPCRInset}.}
    \label{fig:RelTime}
\end{figure}

This ``compression wave'' evolves outwards and interacts with periodic images at the boundary, leading to several oscillations until the difference settles down to zero. The point when $\Gamma_\text{hot}(t)-\Gamma_\text{cold}(t)$ settles down to zero, statistically speaking, can be used to define the thermal relaxation time of the system, $\tau_\text{relax}$, see Fig. \ref{fig:RelTime}. On average, collisions between particles of different speed will tend to equalize their speeds, see Eq. \eqref{Eq:VelUpd}. Thus, \textit{the total collision rate of the system determines the relaxation timescale for velocity inhomogeneities}, and so at lower densities, where collisions are less frequent, relaxation times are longer, and at higher densities, the system relaxes quickly.

\section{Structural Relaxation of Amorphous and Glassy Configurations}
\label{sec:StructRel}
As we mentioned in the introduction, it is known that the HS equation of state bifurcates with increasing density, with a ground state branch corresponding to ``frozen'' crystalline configurations, and with a metastable branch corresponding to  amorphous fluid-like or glass-like configurations. Disordered configurations on the metastable branch have a higher pressure than the ordered configurations on the ground state branch at the same density. Above a certain density on this metastable branch, structural relaxation timescales become extremely long, and the states are reasonably described as ``glassy'' \cite{berthier2011theoretical,speedy1994quench,alder1960studies}. (Another practical definition is offered by Speedy \cite{speedy1997pressure}, where the system is considered glassy if the RMS displacement of any given sphere is less than a sphere diameter over a long simulation run.) We prefer the term ``glassy'', rather than ``glass,'' to avoid the subtleties of the discussion surrounding the distinction between thermodynamic and kinetic glass transitions, see, for example, the discussion in \textcite{parisi2010mean} and the references therein. 

\begin{figure}
    \centering
    \includegraphics[width=1.0\columnwidth]{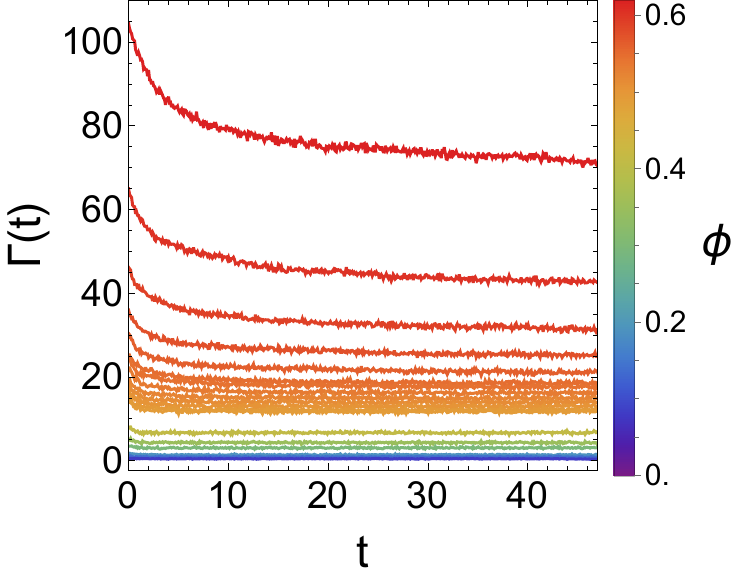}
    \caption{Average single particle collision rate vs. time for a system where the initial positions were produced from a very rapid compression, using the LS procedure with  compression rate $\gamma=1$, thereby placing the high density systems on the metastable, spatially disordered, branch of the phase diagram. The highest density state, $\phi=0.62$, is close to the maximally random jammed density, $\phi_\text{MRJ}\approx0.64$ \cite{torquatorandomclosepacking}, or the closely related random close packing (RCP) density \cite{zaccone2022explicit,zaccone2023,zaccone2025complete}, which is approximately the highest possible density for disordered sphere packings in $d=3$. As in Fig. \ref{fig:FastRel}, an initial density-dependent transient behavior is observed at early times. However, the high density runs do not settle down to equilibrium, instead exhibiting a significantly higher pressure than the equilibrium systems at the same density and energy, see Fig. \ref{fig:SPCRInset}. The same range of densities is used as in Figs. \ref{fig:SPCRInset}, \ref{fig:SPCRThermRel}, and \ref{fig:RelTime}, and once again all simulations are performed at the same total energy.}
    \label{fig:SPCRAmorphRel}
\end{figure}

To examine the non-equilibrium dynamics of hard spheres in response to a different kind of ``perturbation'' to the initial state, we prepare our HS system in disordered states by rapidly compressing the system, using the LS procedure with an extremely rapid growth rate $\gamma=1$ \cite{speedy1994quench}, and then we allow the system to evolve freely under EDMD evolution. Using such a high compression rate, the spheres do not have sufficient time to find a crystallized configuration at high densities, and the system is trapped in a disordered state. The configurations obtained in this manner do not exhibit an ordered spatial structure with well-defined crystal axes, like in Fig. \ref{fig:SPCRInset} and Fig. \ref{fig:FastRel}, and when these states are allowed to relax under EDMD evolution, they exhibit qualitatively different behavior of the single-particle collision rate, see Fig. \ref{fig:SPCRAmorphRel}.

\begin{figure}
    \centering
    \includegraphics[width=1.0\columnwidth]{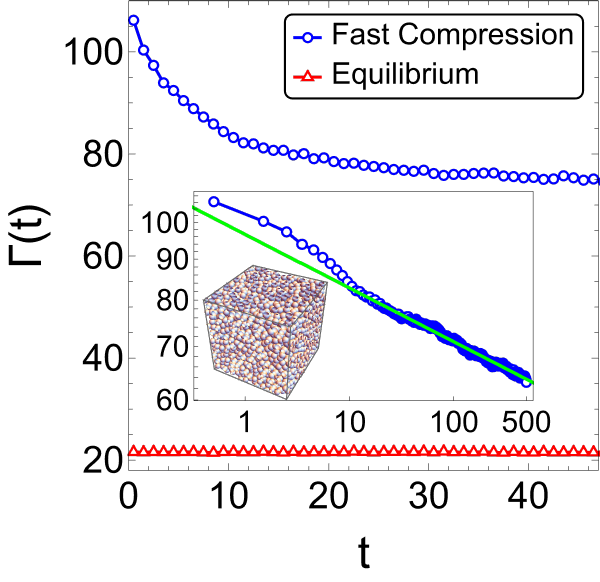}
    \caption{Average single-particle collision rate vs. time for the highest-density glassy system considered in Fig. \ref{fig:SPCRAmorphRel} (blue points), at packing fraction $\phi=0.62$, with the equilibrium result at the same density for comparison (red points). A large difference in collision rate, and hence pressure, is observed between the two systems, although the density and energy are the same. The inset shows the blue curve in log-log scale for a longer duration. The collision rate exhibits a very slow power law decay $\Gamma \sim t^{-0.067}$ after an initial transient. A na\"{i}ve extrapolation would suggest that the structural relaxation time for this inital perturbation would be $\tau\sim\mathcal{O}(10^9)$, dramatically longer than the thermal relaxation times observed previously in Fig. \ref{fig:RelTime}.}
    \label{fig:SlowRel}
\end{figure}

\begin{figure}
    \centering
    \includegraphics[width=1.0\columnwidth]{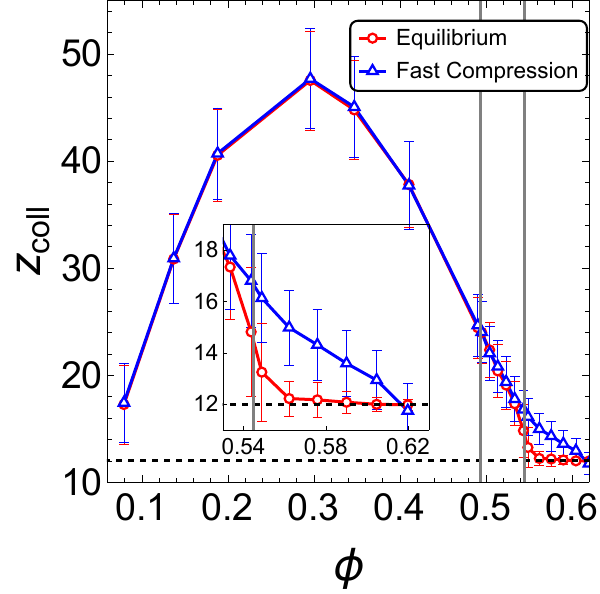}
    \caption{The average ``collisional contact number'' $z_\text{coll}$, or the number of contacts (excluding repeat collisions) encountered by a given particle during time evolution for simulations with a fixed number of particles and energy and run for a fixed duration of simulation time, calculated over a range of densities for both the slow-compression and fast-compression systems, which correspond to crystalline and glassy configurations respectively at high densities. The horizontal dashed line indicates the kissing number for monodisperse spheres in 3D, 12, which would be the expected value of $z_\text{coll}$ for a perfect FCC crystal. We see the effect of rapid compression is only observed in the solid regime above the melting density $\phi_\textrm{melt}\approx0.545$. $z_\text{coll}$ captures the degree of ``mixing'' in the fluid phase, and the degree of structural rearrangement in the solid phase, see discussion in main text. All simulations were run for a total time of $T=50.0$.}
    \label{fig:NumPartners}
\end{figure}

At first glance, Fig. \ref{fig:SPCRAmorphRel} seems to indicate similar behavior to the collision rates observed in Fig. \ref{fig:SPCRThermRel}, although with a slightly different profile for the ``transient.'' However, when plotted alongside the equilibrium curve at the same density (i.e., the crystallized configuration, prepared with the slow growth rate $\gamma = 10^{-5}$ \cite{skoge2006packing}), we see a stark difference, see Fig. \ref{fig:SlowRel}. Although the high-density amorphous system is technically ``relaxing,'' in the sense that the collision rate is decaying towards its equilibrium value with time, the rate of decay is extremely slow, $\Gamma(t)\sim t^{-0.067}$, indicating that the lifetime of this metastable configuration is many orders of magnitude longer than the length of the simulation. (A na\"{i}ve extrapolation of that curve would suggest $\tau>10^9$.) We find that such long-lived metastable states appear between $\phi = 0.534$ and $\phi=0.544$, just below the melting density $\phi_\text{melt}\approx0.545$, although a more recent estimate places the melting density closer to $\phi_\text{melt}\approx0.543$ \cite{smallenburg2024}. At this point, a small gap appears between the collision rate of the rapidly compressed systems and the equilibrium systems at the same density, and this gap widens as the density is increased towards $\phi_\text{MRJ}$, where it diverges. We find that for all densities $\phi>0.534$ that we considered, the system does not fully relax to the equilibrium value of the pressure during the course of the simulation, and hence full structural relaxation to a crystalline configuration does not occur during the simulation. All simulations were run for total time $T=50.0$. 

The existence of these long-lived metastable states in HS systems has been discussed in the literature for at least 65 years, since the pioneering work of Alder and Wainwright \cite{alder1960studies}. The very slow decay behavior of the pressure was also observed by \textcite{nearlyjammed} and \textcite{speedy1997pressure}, although in the former study their jamming protocol used a time-dependent $\gamma$, which was initially fast to suppress crystallization, but was slowed at higher densities to allow for more particle rearrangement and thereby get much closer truly jammed configurations than we do in this study. \textcite{nearlyjammed} refer to this behavior of the pressure as a ``pressure leak,'' and they observe it is always associated with positional rearrangements of the spheres, an observation which we confirm in a different manner below.

We can investigate spatial structure in EDMD simulations of hard sphere systems by calculating the average ``collisional contact number,'' $z_\text{coll}$, defined to be the average number of contacts (excluding repeat collisions) encountered by a given sphere during the course of the simulation, see Fig. \ref{fig:NumPartners}. This quantity was originally introduced by \textcite{nearlyjammed}, who investigated its behavior very close to the jamming point. Remarkably, in this limit, the asymptotic value of $z_\text{coll}$ is $\approx5.7$ \cite{nearlyjammed}, which indicates that the blue curve in Fig. \ref{fig:NumPartners} will continue to decrease sharply as the density is increased to $\phi_\text{MRJ}\approx0.64$. We note that these authors only counted contacts over a restricted time interval of 100 collision per particle, a restriction which we do not impose here in order better characterize positional rearrangements of spheres during the course of the whole simulation. It is simple to obtain $z_\text{coll}$ from the collision sequence in an EDMD simulation (and hence the hard sphere causal graph, see Section \ref{sec:CG}), and so one can easily calculate $z_\text{coll}$ at any point in the phase diagram, even in the fluid phase very far from the jamming point. 

Regardless of the rate of compression for the initial conditions, we observe a peak in the curve of $z_\text{coll}$ vs. density. As all simulations were run for the same amount of time and at the same energy, we can understand this profile as follows. At low densities, collisions are infrequent, and hence the contact number per particle is low. At intermediate densities, collisions are much more frequent between particles, but the fluid is also easily able to mix, leading to a very large number of contacts per particle, with a peak at $\phi\approx0.296$. At high densities, collisions are extremely frequent, but mixing is increasingly suppressed, and the spatial arrangement of spheres mostly remains close to that of the initial state. 

Fig. \ref{fig:NumPartners} corroborates the claim made earlier that the initial condition preparation is irrelevant in the fluid regime, where the two systems prepared with $\gamma=1$ and $\gamma=10^{-5}$ exhibit identical behavior. The difference in compression rate during the LS procedure used to prepare the initial conditions manifests upon solidification though, where the crystalline systems essentially fluctuate about a regular packing, which is revealed by the convergence of $z_\text{coll}$ to the kissing number, 12, for 3D monodisperse spheres, the expected result for a perfect close-packed FCC crystal where in the limit no particle rearrangements are possible. In the amorphous system in the solid regime, by contrast, particles encounter more neighbors at the same density under time evolution. This is an unexpected finding, as we know that asymptotically $z_\text{coll}\rightarrow 5.7$, significantly less than $z_\text{coll}=12$ observed for crystalline configurations. This low asymptotic value of $z_\text{coll}$ stems from the fact that disordered sphere packings are characterized by a complex and spatially inhomogeneous network
of near-contacts between neighboring spheres, where most spheres are tightly pinned by a
few neighbors, typically fewer than the kissing number (which explains why $\phi_\text{MRJ}<\phi_\text{FCC}$). And so $z_\text{coll}>12$ in the glassy regime up to $\phi\approx0.62$ implies that there must be positional rearrangement of spheres in the glassy systems under time evolution, despite the tightly ``pinned'' local configurations of spheres, see Section \ref{sec:CGForGlassyStates} and the figures therein. This result of positional rearrangement of spheres in the glassy systems matches claims made by \textcite{nearlyjammed}, and indicates that \textit{these high density amorphous states retain some fluid-like characteristics and are not ``true'' glasses, as perfect caging \cite{charbonneau2017glass,parisi2020theory} is incompatible with structural rearrangement of the spheres over time}. 

\section{Causal Graph Analysis}
\label{sec:CG}

At moderate densities, the effect of collisions in hard sphere systems can no longer be treated as a small perturbation on top of an ideal gas or a perfect FCC crystal, and so direct numerical simulation is the primary technique for determining how the interactions between particles dictate the global structure and dynamics of HS systems in this regime. Once one has numerical simulation data, generally in the form of positions and/or velocities of the particles, a quantitative description of structure and dynamics is primarily obtained in the literature through the use of various correlation functions, such as the radial distribution function or the closely related static structure factor, which provide statistical descriptions of the distribution of particles in space and/or time \cite{hansen2013theory}. Alternatively, as we just saw, one can consider single-particle aspects such as the collision rate or the collisional contact number, which also provide some useful information. Another very general approach for investigating the effects of interactions between constituent parts of a system on its global structure and behavior, which is ubiquitous in the scientific literature, is to use graphs/networks \cite{newman2018networks,Wolfram2002}. In this section and the following, we introduce a network-based approach for characterizing hard sphere systems, and we show how the results presented in the previous sections can also be interpreted naturally in terms of our network construction. We stress that network analysis is complementary to traditional phase space analysis for these systems, as each offers a different quantitative perspective on the same underlying physics.

There are a number of possible ways to define a network structure from a hard sphere system. We could treat particles as nodes, with collisions (or proximity in space) forming edges between them; this construction is ubiquitous in the context of jamming studies, and goes by the name of the ``contact network,'' see, for example, \cite{donev2004jamming,van2009jamming,torquato2010jammed}. While very useful in the context of jamming and for characterizing the static properties of hard sphere solids, these contact networks are not well suited to dynamical calculations, and are generally restricted to high-density configurations, often at infinite pressure. Alternatively, we could treat pairs of particles as nodes, representing all possible two-particle collisions, and as the EDMD algorithm steps from collision to collision, these nodes could be connected with directed edges. In this case, EDMD evolution is represented in terms of a walk on this graph. Although this construction captures dynamics in some manner, it obscures spatial relationships between particles as well as the details of single-particle motion, and we are presently unaware of any application of such a construction. We introduce here a third approach, which preserves some information pertaining to spatial relationships between particles, like in contact networks, while simultaneously allowing for the description of dynamics and preserving certain aspects of single-particle motion. We call this construction the \textit{hard sphere causal graph}. 

\begin{figure}
    \centering
    \includegraphics[width=1.0\columnwidth]{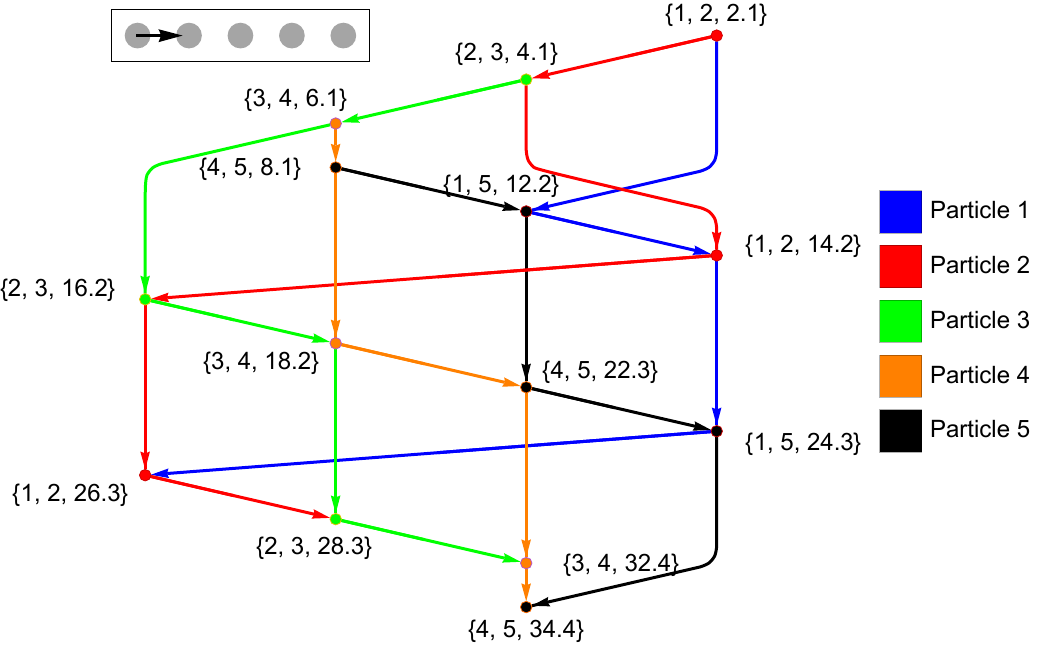}
    
    \caption{An example of a hard sphere causal graph, a directed acyclic graph (DAG) constructed from the time-evolution of the hard sphere system. Here, we consider the simple case of a ``Newton's cradle'' configuration of five particles arranged in a line with periodic boundary conditions (see inset). Initially, only the particle on the left is moving, and the momentum is transferred down the line three times. Edges correspond to individual particles, see detailed explanation in the main text, and so each particle is assigned a unique color to help the reader interpret the graph. Nodes are labeled by the pair of particles involved in the collision, as well as the time at which the collision takes place: $(p_i,p_j,t)$. \textit{We stress that the causal graph is not a spacetime diagram, and so the layout of the graph on the page is arbitrary}; instead, what is important is the connectivity structure (and the node labels). Nevertheless, we can always choose a layout such that time increases as one moves down the page \cite{newman2018networks}, which we do here. }\label{fig:colorCG}
\end{figure}

In the hard sphere causal graph, each collision in the EDMD evolution is represented as a separate node in the graph.  To distinguish between repeated collisions between the same pair of particles, we label nodes with a triple of numbers, the identity (i.e., index) of each particle, and the time of the collision: $(p_i,p_j,t_\text{coll})$, where the order of the first two entries is unimportant. We construct the edges as follows. During EDMD evolution, each particle $i$ undergoes a sequence of collisions with other particles (arranged in increasing order of time), which are represented as a sequence of nodes $c^{(i)}_1,c^{(i)}_2,c^{(i)}_3,c^{(i)}_4,...$ in the graph; we choose to connect these nodes by directed edges according to the rule $c^{(i)}_1\rightarrow c^{(i)}_2,c^{(i)}_2\rightarrow c^{(i)}_3,c^{(i)}_3\rightarrow c^{(i)}_4,...$ and so on. Evidently, we can think of each of these edges as ``following'' particle $i$ as it evolves from one collision to the next. Thus, in the hard sphere causal graph, each edge should be interpreted as referring to a specific particle, and by following the edges associated with a given particle through the graph, the collision history of that particle can be recovered.  We construct such collision sequences/single-particle paths for all particles $i$, and together they form the causal graph, which will be a directed acyclic graph (DAG), as all directed edges connect a collision which is earlier in time to a collision which is later in time, and hence directed cycles will be forbidden. And since each collision $(p_i,p_j,t_\text{coll})$ involves two particles, each node will form an intersection between two paths, and so the causal graph will be a highly interconnected union of the single-particle paths, see Fig. \ref{fig:colorCG} for a simple example, where the single-particle paths have each been assigned a unique color to help the reader interpret the graph.

\begin{figure}
    \centering
    \includegraphics[width=1.0\columnwidth]{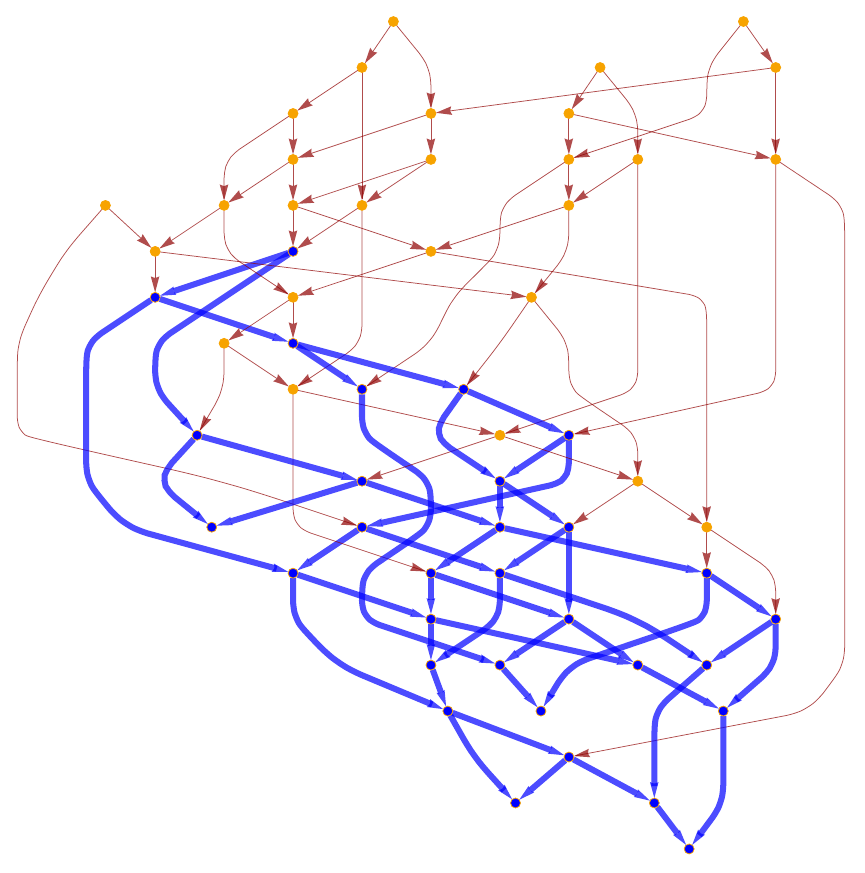}
    \caption{Hard sphere causal graph for a short evolution of a system of 15 spheres in the fluid phase in three dimensions, with a typical ``out-component'' highlighted in blue. The out-component is constructed by following all edges out from a given node, in this case until the boundary of the graph is reached. Once again, we choose to show time increasing as one moves down the page, but the layout of the graph on the page is ultimately arbitrary.}
    \label{fig:CGFLC}
\end{figure}

Causal graphs, in particular, have been investigated in the context of discrete models of complex systems and fundamental physics, derived from state rewriting rules \cite{Wolfram2002,wolfram2020class}. These structures have their origins in causal set theory, where a causal set is a partially ordered set consisting of a collection of discrete spacetime events with causal relations between events \cite{kronheimer1967structure, surya2019causal}. The key point emphasized in \cite{Wolfram2002, wolfram2020class, arsiwalla2024pregeometric, arsiwalla2021homotopies,wolfram2023secondlaw} is that causal graphs serve as foundational data structures of any abstract computational system, one consisting of discrete state transformations. These graphs encapsulate the state space of the system, together with causal relations between states (comparable to traditional phase space information, plus causal structure). Hence, causal graphs are useful not only for analyzing models pertaining to discrete spacetime, but also for computational descriptions of many-body phenomena, such as the EDMD simulations of hard sphere systems that we undertake here. 

The advantages associated with this new causal graph construction may not be immediately obvious, so we now list a few of them here. Firstly, the causal graph is invariant under a large group of (global) transformations to the underlying system of spheres. Since it depends only on the collision sequence for the HS system, one will construct the same causal graph if the system is rotated, translated, Galilean boosted, reflected, or time-translated. This is a basic consequence of Newton's laws of motion.

Secondly, the causal graph is easy to compute and efficient to fully store in computer memory, requiring just two \verb|int|-type (integer) variables and one \verb|double|-type (floating point) variable per collision to save, regardless of the spatial dimension $d$ in which one simulates the hard spheres. Thus, it is possible to save the entire causal graph, even for simulations with billions of collisions, without requiring enormous amounts of memory (tens of gigabytes at most in the case of extremely long simulations). Once the collision list has been saved, the connectivity structure (i.e., adjacency matrix) of the causal graph can be easily constructed as a postprocessing step using the procedure described above, and for large graphs, all single-particle paths can be constructed from the collision list in parallel and without communication for additional speedup. We note also that it is very straightforward to augment existing HS molecular dynamics codes to record this collision sequence from the which the causal graph is constructed. 

Thirdly and most importantly, the HS causal graph can be used calculate, from a single underlying structure, several physical quantities of interest, which is especially appealing given the two previous points. There are two natural quantities which appear when considering the causal graph construction: paths, see Fig. \ref{fig:colorCG}, and out-components \cite{newman2018networks}, see Fig. \ref{fig:CGFLC}. We shall show that both quantities can be used to calculate relevant properties of the HS system.

The path associated with a given particle $k$ is a sequence of collisions, where, as we saw, each collision is labeled generically as $(p_i,p_j,t_\text{coll})$, and where one of the first two numbers is always $k$. From such a sequence, one can easily construct two useful quantities. The sequence of \textit{collision times} can be used to construct the single particle collision rate, $\Gamma(t)$, which is related to the system pressure and hence the equation of state via an affine transformation \cite{alder1960studies,erpenbeck1984molecular,hoover1967studies,engel2013hard}; $\Gamma(t)$ can also be used as a simple characterization of the non-equilibrium dynamics of the system, see Sections \ref{sec:ThermRel} and \ref{sec:StructRel}. The other useful quantity that can be calculated from a single-particle path is the ``collisional contact number,'' $z_\text{coll}$, \cite{nearlyjammed} defined to be the number of other particles (not counting repeat collisions) encountered by a given particle during the simulation. As we already saw in Section \ref{sec:StructRel}, the collisional contact number enables us to quantify the degree of structural rearrangement of spheres at high densities under time evolution and the degree of ``mixing'' of the fluid at lower densities.
\footnote{Another quantity which one can calculate from single-particle paths, and which can be used to characterize the dynamics of spatial mixing and structural rearrangement in more detail, is the number of collisions between particles $i$ and $j$ between times $t$ and $t'$, $z_{ij}(t,t')$. This is a (time-dependent) matrix which captures in great detail the collision history of the system during a given time interval, and we shall not investigate it any further in this study. }
\textit{Evidently, all the results presented in Sections \ref{sec:ThermRel} and \ref{sec:StructRel} and in Figs. \ref{fig:SPCRInset}-\ref{fig:NumPartners} can alternatively be interpreted as coming from analysis of single-particle paths in the hard sphere causal graph.} We note that both $\Gamma(t)$ and $z_\text{coll}$ can, of course, easily be calculated without the causal graph as well, as they only utilize a subset of the connectivity available in the causal graph, but they are nonetheless natural to define, from a single underlying structure, and easy to calculate once one has the causal graph. 

Next, we turn our attention to causal graph out-components, defined as the subgraphs encountered when following all directed edges out from a given node for a certain number of steps or for a given amount of time \cite{newman2018networks}, see Fig. \ref{fig:CGFLC}. In Mathematica/Wolfram Language,  the out-component can be easily constructed from a DAG using the ``VertexOutComponentGraph'' function \cite{VertexOutComponentGraph}. These out-components allow one to visualize the ``causal influence'' of any given collision on the rest of the particles in the system over time, by progressively tracking all particles which are ``causally connected'' (through collisions) to the initial pair of spheres. (The authors of \cite{collisioncascade} called this growing set of causally connected spheres a ``collision cascade,'' but they assumed all other particles were initially stationary, a restriction we do not impose here.) We shall use causal graph out-components to explore the relationship between spatial disorder and enhanced pressure for hard sphere systems in the glassy regime in the next section, Section \ref{sec:CGForGlassyStates}. In contrast to the single-particle paths, the out-components do utilize the full connectivity structure of the causal graph, and are therefore not simple to define for a HS system without constructing the causal graph or an equivalent structure.

Before concluding this section, we offer two brief remarks about implementation. Firstly, in the case of very long simulations, which are sometimes required for simulating glassy systems in particular, the causal graphs can become very large, since they contain as many nodes as there are collisions, and twice as many edges. (Our longest high-density simulation for this study consisted of 650 million collisions, and so the corresponding causal graph had 650 million nodes and 1.3 billion directed edges.) Representation of DAGs with this many nodes and edges can be prohibitively expensive and/or slow when using out-of-the-box tools, as can performing computations on these graphs. However, one can handle such large causal graphs easily by using an ``edge list'' representation of the adjacency matrix, which works as follows. Since the collisions can be uniquely sorted by time, we can assign to each collision a unique integer, $c$. Then directed edges can be represented as ordered pairs, $(c_\text{start},c_\text{end})$. The adjacency matrix may then be stored very compactly as an $N_\text{edge}\times2$ matrix of \verb|int|-type variables. The second important remark is that the node labels must be retained in order to extract useful information from the causal graph; in this sense, the causal graph is not quite as ``parsimonious'' as, say, the static contact network, for which only ``topological'' connectivity information, such as the node degree distribution, is required to extract useful physical information.

\section{Investigation of High-Density States Using the Causal Graph}
\label{sec:CGForGlassyStates}

\input{LightConeGrowth}

\begin{figure}
    \centering
    \includegraphics[width=1.0\columnwidth]{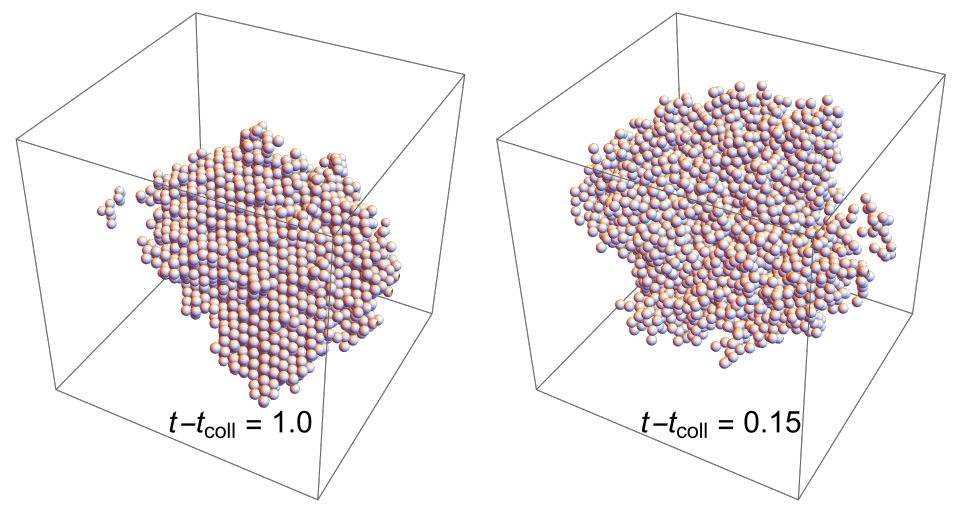}
    \caption{Visualization of spheres in a typical causal graph out-component, same as Fig. \ref{fig:LightConeGrowth}, at high density for the ordered  (left) and disordered (right) configurations for larger simulations of $N=32^3$ particles at packing fraction $\phi=0.62$ and simulated for a longer time. The out-component for the crystalline configuration contains 5929 particles, and the glassy one contains 6835 particles. One can see the spontaneously chosen crystal axes in the left figure. Note that spheres which appear to be ``floating'' disconnected from the main structure have in fact simply passed through the periodic boundary.}
    \label{fig:FLCSpheres}
\end{figure}

As we demonstrated in Section \ref{sec:StructRel}, the configurations on the metastable branch exhibit a higher collision rate and hence pressure compared to those on the ground state branch at the same density. This enhanced collision rate must be a consequence of the spatial disorder in the metastable systems, and the causal graph allows us to see in detail why this is the case. In particular, recalling Fig. \ref{fig:CGFLC}, we can visualize, in space this time, the growth of some typical causal graph out-components for these simulations, see Figs. \ref{fig:LightConeGrowth} and \ref{fig:FLCSpheres}. These out-components allow one to visualize the ``causal influence'' of any given collision on the rest of the particles in the system, by progressively tracking all particles which are ``causally connected'' (through collisions) to the initial pair of spheres. Given enough time, collisions will propagate through the entire system, progressively ``infecting'' more spheres, and the out-component will contain all the particles; at this point, it does not provide any valuable information (apart from the total collision rate, which is asymptotically captured in the number of new nodes added to the causal graph per unit time). However, before this time, growth of the out-component reveals the spatial structure of the material by illuminating the network of near-contacts between spheres. (Recall that simultaneous collisions do not occur in EDMD in the absence of symmetry, so simultaneous contact of three or more spheres will never occur in one of these simulations. By ``near-contact'' we mean spheres which were previously in contact in the sense of having collided, and are therefore probably nearby in space.) Nevertheless, the spatial structures produced from growing a causal graph out-component from a given node/collision may serve as the starting point for constructing a finite-pressure analog of the contact networks extensively used in jamming studies of hard spheres, see, for example, \cite{donev2004jamming,van2009jamming,torquato2010jammed}. 

In Fig. \ref{fig:LightConeGrowth}, we observe that collisions propagate through both crystalline and glassy systems in an approximately spherical ``cloud'' surrounding any given collision, an observation similar to that made in \cite{collisioncascade}, but that they move through the disordered system at roughly six or seven times the rate they do through the crystal (for $\phi = 0.62$). This difference originates with the fact that, although the total density of both systems is the same, the local densities are not necessarily so, and the disordered system has a complex and spatially inhomogeneous network of near-contacts between neighboring spheres, characteristic of jammed configurations and high density disordered sphere packings \cite{torquatorandomclosepacking,donev2004jamming,van2009jamming,torquato2010jammed}, but at finite pressure. In a glassy state, most spheres are tightly pinned by a few neighbors, typically fewer than the kissing number for 3D monodisperse spheres (i.e., 12) \cite{conway2013sphere}, and hence a collision ``cascade'' can propagate quickly through such a configuration, analogous to momentum propagating rapidly through the interior beads of a  Newton's cradle. This process is the origin of the ``spindly'' structures in the lower row of Fig. \ref{fig:LightConeGrowth}. By contrast, the crystal at the same density fluctuates around a regular lattice configuration, where hardly any such local ``pinning'' occurs, and hence the out-component grows in space in a much more isotropic manner. We defer the further quantitative investigation of the growth of these structures in space until a future study, as very large systems (and hence causal graphs) are required to address this question without interference from periodic boundary effects, and at present, such a study would be very expensive computationally.

\section{Conclusions}
\label{sec:conclusion}
In summary, we have performed event-driven molecular dynamics simulations to investigate the non-equilibrium dynamics of hard spheres in response to two classes of perturbations. Firstly, we drove the velocity distribution out of equilibrium by inhomogeneously heating the system in a central region. We found that relaxation dynamics are comparatively fast and are governed by the total collision rate of the system, as the effect of individual collisions is on average to equalize the speeds of the particles, see Eq. \eqref{Eq:VelUpd}, irrespective of the spatial structure of the system. The collision rate dependence of the relaxation time leads to an approximately inverse relationship between the thermal relaxation time and the total system density. Furthermore, relaxation to equilibrium is preceded by a radially expanding longitudinal ``compression wave'' which moves through the material and is eventually dissipated by particle-particle collisions and the periodic boundary. 

Secondly, we drove the spatial distribution out of equilibrium by performing a rapid compression with the Lubachevsky-Stillinger procedure, thereby placing the high density systems on the metastable branch of the equation of state. We found that structural relaxation of these configurations is strongly suppressed above a certain density, and indeed never occurs within the simulation times considered in this study. The pressure of such metastable states exceeds that of their equilibrium counterparts at the same density, and this difference widens as the maximally random jamming density is approached, at which point the pressure of the metastable system diverges. Although these metastable systems are unable to fully relax to the equilibrium crystalline structure, we found that structural rearrangement is nevertheless still occurring at some level. 

It follows from our first perturbation studies in Sec. \ref{sec:ThermRel} that the enhanced collision rate in glassy systems will lead to extremely rapid \textit{thermal} relaxation of these systems, far faster than in a crystalline system at the same density, although full  relaxation to equilibrium requires structural relaxation, which only occurs over very long timescales, if at all. This observation, that hard sphere glasses restore equipartition extremely quickly and hence have a relatively high thermal conductivity, extends the validity of the observation that glasses, although necessarily out of equilibrium, share many features in common with equilibrium systems, and hence can be treated under certain circumstances with equilibrium methods, see, for example, \cite{parisi2010mean,parisi2020theory} and the references therein. One can plausibly argue, based on the results we have presented here, that perturbed hard sphere glasses will also relax quickly to ``equilibrium,'' and hence we speculate that some techniques and approximations which assume small deviations from equilibrium are likely to be reliable for describing HS systems in this regime. 

We have also introduced the hard sphere causal graph, a directed acyclic graph constructed from the dynamical evolution of the hard sphere system. The causal graph naturally encapsulates several quantities which have previously been used in the literature to analyze hard sphere systems, including the single-particle collision rate, the collisional contact number \cite{nearlyjammed}, and the notion of a collision cascade \cite{collisioncascade}. The causal graph offers a memory-efficient summary of the dynamical history of the system, requiring just two \verb|int|-type variables and one \verb|double|-type variable per collision to store, regardless of spatial dimension of the system. We have argued that the causal graph encapsulates several useful metrics for characterizing HS systems, and therefore serves as a useful complement to typical phase space analysis of these systems. Individual particles manifest as (directed) paths in the causal graph, from which can be extracted the single-particle collision rate $\Gamma(t)$, which is a simple measure of both equilibrium and non-equilibrium dynamics of the system, as well as the ``collisional contact number,'' $z_\text{coll}$, which can be used to quantify the degree of mixing in fluid configurations and the degree of structural rearrangement (or lack thereof) in solid configurations. It is also very natural from the perspective of the causal graph to define causal graph out-components, which enable one to directly visualize the spatial structure of the material, and which provide a principled starting point for extending the widely used notion of a contact network to systems at finite pressure and at densities far from the jamming point. Causal graphs can be easily incorporated into existing molecular dynamics codes, and we suggest that they may at least be useful for the analysis of other hard particle systems, such as \cite{van2009jamming,torquato2010jammed}. Presumably, causal graphs may prove useful for the analysis of other many-body systems.

We suggest, in no particular order, several additional possibilities for future investigation based on the considerations presented in this study.  1.) What is the relationship between the hard sphere causal graph and the various HS transport coefficients \cite{transportcoeffs}, as well as the speed of sound, in the fluid, crystalline, and glassy regimes? 2.) What are the precise growth characteristics (e.g., growth rate, fractal dimension, etc.) of the out-components, considered in space, in the various high-density regimes? 3.) How can the causal graph be defined for many-body systems with smooth potentials for which collisions are not instantaneous, for example soft spheres or Lennard-Jones fluids, and do these systems show any qualitative differences when compared with hard spheres at the level of the causal graph? 
4.) What can the tools and techniques employed in this study tell us about more general non-equilibrium processes involving hard spheres?

We shall elaborate briefly on this last question. The perturbation studies considered Sections \ref{sec:ThermRel} and \ref{sec:StructRel} of this work focus on relaxation phenomena, which constitute a narrow subset of the broad class of possible non-equilibrium processes in HS systems. All simulations in this study are performed in the microcanonical ensemble, with fixed energy, particle number, and system volume, subject to no external forces, and where the non-equilibrium dynamics follow entirely from an initial state prepared out of equilibrium. Despite the significant restrictions in this particular study, the tools used here (e.g., the collision rate, the collisional contact number, and the causal graph) were specifically chosen because they are very general and can be straightforwardly applied to analyze a large variety of more ``exotic'' non-equilibrium processes in HS systems. The simplest example of such an extension would be to apply these tools to the Lubachevsky-Stillinger (LS) procedure itself (see Sec. \ref{sec:EDMD}), in which case the system is no longer in the microcanonical ensemble but is rather entirely out of equilibrium and is being “driven” by energy being supplied to the system through the inelastic collisions between particles. Other non-equilibrium scenarios might include subjecting the spheres to a (possibly time-dependent) external potential (e.g., the ``Boltzmann Breather'' configuration \cite{BoltzmannBreather}), adding fluctuations and/or dissipation \cite{zwanzig2001nonequilibrium} to the dynamics, or having a net ``flux'' of particles through the system. The only ingredients required to use the tools advocated in this study are some sort of dynamical evolution procedure and a well-defined notion of collision between the spheres.

Finally, in our investigation of HS systems, we have considered a complementary network perspective, based on causality relations between components of a many-body system. This suggests a new avenue of exploration for discrete gravity approaches, including causal set theory, in which the thermodynamic investigation of causal graphs may help realize a dual description of spacetime geometry.

\section*{Acknowledgements} 

We would like to thank Stephen Wolfram for suggesting the hard sphere causal graph for investigation. We would also like to thank Christopher Wolfram, as well as the rest of the 2020 Wolfram Summer School team, for their valuable contributions to the early exploration of these ideas.  MK would like to thank Aurel Bulgac and Lukasz Fidkowski  for several fruitful discussions. Finally, MK would like to thank Lea Anne Copenhefer and Scott Kafker for their generous contribution of computing resources.

\bibliography{biblio}

\end{document}

%% file: LightConeGrowth.tex
\begin{figure*}[!htb]
    \centering
    \includegraphics[width=\textwidth]{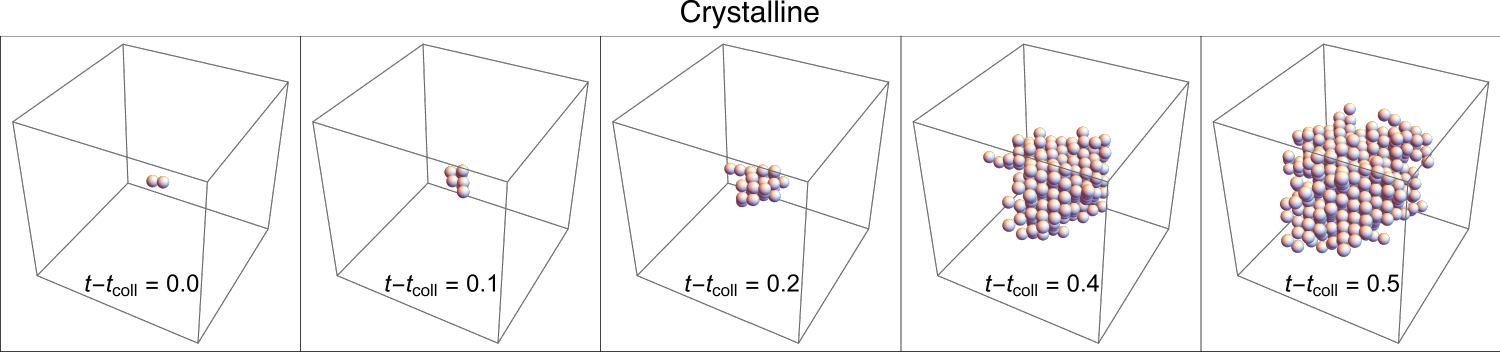}
    \includegraphics[width=\textwidth]{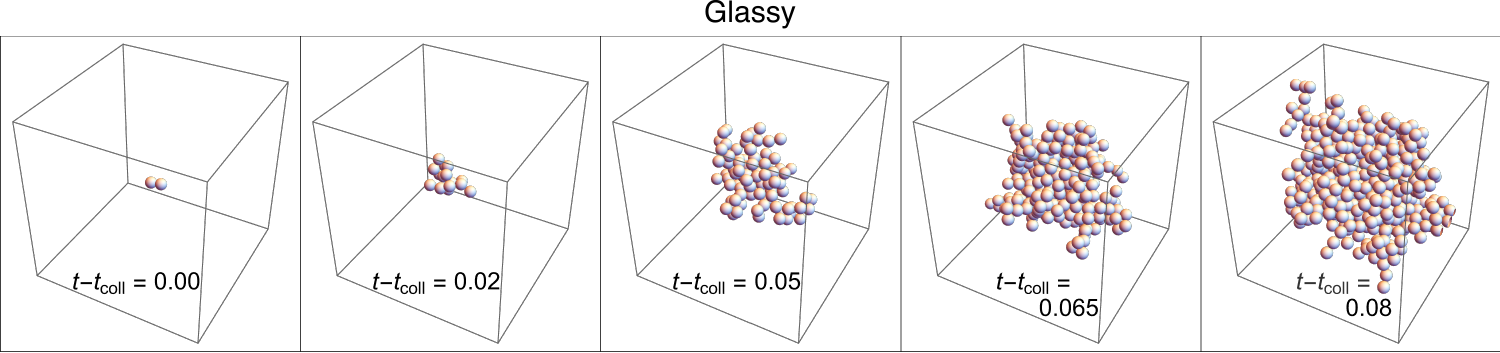}
    \caption{Spatial visualization of the growth of a causal graph out-component for a typical collision, see Fig. \ref{fig:CGFLC}, as a function of time, for a crystalline (top) and glassy (bottom) system at the same density, $\phi=0.62$, and energy. Each out-component, which can also be thought of as a ``collision cascade'' \cite{collisioncascade}, grows from an initial ``seed'' collision involving only two particles (leftmost frame), and gradually encompasses more particles over time. The other particles, which are not yet included in the out-component, are not shown, but the reader should remember that they are present; this figure is showing a subset of the particles associated with a particular out-component, where the full simulations look like the insets in Figs. \ref{fig:FastRel} and \ref{fig:SlowRel} for the crystalline and glassy configurations, respectively. At any one time, the out-component will contain a set of  spheres which are nearly in contact, but not quite (as simultaneous collisions in EDMD occur with probability zero). These structures may be used as the starting point for constructing finite-pressure analogs of contact-networks used in jamming studies of hard spheres, see, for example, \cite{donev2004jamming,van2009jamming,torquato2010jammed}. The out-component encompasses a roughly-spherical, growing collection of spheres as time elapses, centered roughly around the initial ``seed'' collision. However, the glassy phase out-component has a more ``spindly'' structure, where the ``tendrils'' arise from collisions propagating rapidly through locally ``pinned'' configurations in the amorphous geometry. Note also that the out-component in the glassy case grows much faster than the crystalline case.}
    \label{fig:LightConeGrowth}
\end{figure*}

%% file: main.bbl
\begin{thebibliography}{49}%
\makeatletter
\providecommand \@ifxundefined [1]{%
 \@ifx{#1\undefined}
}%
\providecommand \@ifnum [1]{%
 \ifnum #1\expandafter \@firstoftwo
 \else \expandafter \@secondoftwo
 \fi
}%
\providecommand \@ifx [1]{%
 \ifx #1\expandafter \@firstoftwo
 \else \expandafter \@secondoftwo
 \fi
}%
\providecommand \natexlab [1]{#1}%
\providecommand \enquote  [1]{``#1''}%
\providecommand \bibnamefont  [1]{#1}%
\providecommand \bibfnamefont [1]{#1}%
\providecommand \citenamefont [1]{#1}%
\providecommand \href@noop [0]{\@secondoftwo}%
\providecommand \href [0]{\begingroup \@sanitize@url \@href}%
\providecommand \@href[1]{\@@startlink{#1}\@@href}%
\providecommand \@@href[1]{\endgroup#1\@@endlink}%
\providecommand \@sanitize@url [0]{\catcode `\\12\catcode `\$12\catcode
  `\&12\catcode `\#12\catcode `\^12\catcode `\_12\catcode `\%12\relax}%
\providecommand \@@startlink[1]{}%
\providecommand \@@endlink[0]{}%
\providecommand \url  [0]{\begingroup\@sanitize@url \@url }%
\providecommand \@url [1]{\endgroup\@href {#1}{\urlprefix }}%
\providecommand \urlprefix  [0]{URL }%
\providecommand \Eprint [0]{\href }%
\providecommand \doibase [0]{http://dx.doi.org/}%
\providecommand \selectlanguage [0]{\@gobble}%
\providecommand \bibinfo  [0]{\@secondoftwo}%
\providecommand \bibfield  [0]{\@secondoftwo}%
\providecommand \translation [1]{[#1]}%
\providecommand \BibitemOpen [0]{}%
\providecommand \bibitemStop [0]{}%
\providecommand \bibitemNoStop [0]{.\EOS\space}%
\providecommand \EOS [0]{\spacefactor3000\relax}%
\providecommand \BibitemShut  [1]{\csname bibitem#1\endcsname}%
\let\auto@bib@innerbib\@empty
\bibitem [{\citenamefont {Mulero}(2008)}]{mulero2008theory}%
  \BibitemOpen
  \bibfield  {author} {\bibinfo {author} {\bibfnamefont {{\'A}ngel}\
  \bibnamefont {Mulero}},\ }\href {\doibase
  https://doi.org/10.1007/978-3-540-78767-9} {\emph {\bibinfo {title} {Theory
  and simulation of hard-sphere fluids and related systems}}},\ Vol.\ \bibinfo
  {volume} {753}\ (\bibinfo  {publisher} {Springer},\ \bibinfo {year}
  {2008})\BibitemShut {NoStop}%
\bibitem [{\citenamefont {Alder}\ and\ \citenamefont
  {Wainwright}(1957)}]{alder1957phase}%
  \BibitemOpen
  \bibfield  {author} {\bibinfo {author} {\bibfnamefont {Berni~Julian}\
  \bibnamefont {Alder}}\ and\ \bibinfo {author} {\bibfnamefont
  {Thomas~Everett}\ \bibnamefont {Wainwright}},\ }\bibfield  {title} {\enquote
  {\bibinfo {title} {Phase transition for a hard sphere system},}\ }\href
  {\doibase https://doi.org/10.1063/1.1743957} {\bibfield  {journal} {\bibinfo
  {journal} {The Journal of chemical physics}\ }\textbf {\bibinfo {volume}
  {27}},\ \bibinfo {pages} {1208--1209} (\bibinfo {year} {1957})}\BibitemShut
  {NoStop}%
\bibitem [{\citenamefont {Alder}\ and\ \citenamefont
  {Wainwright}(1959)}]{alder1959studies1}%
  \BibitemOpen
  \bibfield  {author} {\bibinfo {author} {\bibfnamefont {Berni~J}\ \bibnamefont
  {Alder}}\ and\ \bibinfo {author} {\bibfnamefont {Thomas~Everett}\
  \bibnamefont {Wainwright}},\ }\bibfield  {title} {\enquote {\bibinfo {title}
  {Studies in molecular dynamics. i. general method},}\ }\href {\doibase
  https://doi.org/10.1063/1.1730376} {\bibfield  {journal} {\bibinfo  {journal}
  {The Journal of Chemical Physics}\ }\textbf {\bibinfo {volume} {31}},\
  \bibinfo {pages} {459--466} (\bibinfo {year} {1959})}\BibitemShut {NoStop}%
\bibitem [{\citenamefont {Alder}\ and\ \citenamefont
  {Wainwright}(1960)}]{alder1960studies}%
  \BibitemOpen
  \bibfield  {author} {\bibinfo {author} {\bibfnamefont {Berni~Julian}\
  \bibnamefont {Alder}}\ and\ \bibinfo {author} {\bibfnamefont
  {Thomas~Everett}\ \bibnamefont {Wainwright}},\ }\bibfield  {title} {\enquote
  {\bibinfo {title} {Studies in molecular dynamics. ii. behavior of a small
  number of elastic spheres},}\ }\href {\doibase
  https://doi.org/10.1063/1.1731425} {\bibfield  {journal} {\bibinfo  {journal}
  {The Journal of Chemical Physics}\ }\textbf {\bibinfo {volume} {33}},\
  \bibinfo {pages} {1439--1451} (\bibinfo {year} {1960})}\BibitemShut {NoStop}%
\bibitem [{\citenamefont {Parisi}\ and\ \citenamefont
  {Zamponi}(2010)}]{parisi2010mean}%
  \BibitemOpen
  \bibfield  {author} {\bibinfo {author} {\bibfnamefont {Giorgio}\ \bibnamefont
  {Parisi}}\ and\ \bibinfo {author} {\bibfnamefont {Francesco}\ \bibnamefont
  {Zamponi}},\ }\bibfield  {title} {\enquote {\bibinfo {title} {Mean-field
  theory of hard sphere glasses and jamming},}\ }\href {\doibase
  https://doi.org/10.1103/RevModPhys.82.789} {\bibfield  {journal} {\bibinfo
  {journal} {Reviews of Modern Physics}\ }\textbf {\bibinfo {volume} {82}},\
  \bibinfo {pages} {789--845} (\bibinfo {year} {2010})}\BibitemShut {NoStop}%
\bibitem [{\citenamefont {Charbonneau}\ \emph {et~al.}(2017)\citenamefont
  {Charbonneau}, \citenamefont {Kurchan}, \citenamefont {Parisi}, \citenamefont
  {Urbani},\ and\ \citenamefont {Zamponi}}]{charbonneau2017glass}%
  \BibitemOpen
  \bibfield  {author} {\bibinfo {author} {\bibfnamefont {Patrick}\ \bibnamefont
  {Charbonneau}}, \bibinfo {author} {\bibfnamefont {Jorge}\ \bibnamefont
  {Kurchan}}, \bibinfo {author} {\bibfnamefont {Giorgio}\ \bibnamefont
  {Parisi}}, \bibinfo {author} {\bibfnamefont {Pierfrancesco}\ \bibnamefont
  {Urbani}}, \ and\ \bibinfo {author} {\bibfnamefont {Francesco}\ \bibnamefont
  {Zamponi}},\ }\bibfield  {title} {\enquote {\bibinfo {title} {Glass and
  jamming transitions: From exact results to finite-dimensional
  descriptions},}\ }\href {\doibase
  https://doi.org/10.1146/annurev-conmatphys-031016-025334} {\bibfield
  {journal} {\bibinfo  {journal} {Annual Review of Condensed Matter Physics}\
  }\textbf {\bibinfo {volume} {8}},\ \bibinfo {pages} {265--288} (\bibinfo
  {year} {2017})}\BibitemShut {NoStop}%
\bibitem [{\citenamefont {Parisi}\ \emph {et~al.}(2020)\citenamefont {Parisi},
  \citenamefont {Urbani},\ and\ \citenamefont {Zamponi}}]{parisi2020theory}%
  \BibitemOpen
  \bibfield  {author} {\bibinfo {author} {\bibfnamefont {Giorgio}\ \bibnamefont
  {Parisi}}, \bibinfo {author} {\bibfnamefont {Pierfrancesco}\ \bibnamefont
  {Urbani}}, \ and\ \bibinfo {author} {\bibfnamefont {Francesco}\ \bibnamefont
  {Zamponi}},\ }\href {\doibase 10.1017/9781108120494} {\emph {\bibinfo {title}
  {Theory of simple glasses: exact solutions in infinite dimensions}}}\
  (\bibinfo  {publisher} {Cambridge University Press},\ \bibinfo {year}
  {2020})\BibitemShut {NoStop}%
\bibitem [{\citenamefont {Berthier}\ and\ \citenamefont
  {Biroli}(2011)}]{berthier2011theoretical}%
  \BibitemOpen
  \bibfield  {author} {\bibinfo {author} {\bibfnamefont {Ludovic}\ \bibnamefont
  {Berthier}}\ and\ \bibinfo {author} {\bibfnamefont {Giulio}\ \bibnamefont
  {Biroli}},\ }\bibfield  {title} {\enquote {\bibinfo {title} {Theoretical
  perspective on the glass transition and amorphous materials},}\ }\href
  {\doibase https://doi.org/10.1103/RevModPhys.83.587} {\bibfield  {journal}
  {\bibinfo  {journal} {Reviews of modern physics}\ }\textbf {\bibinfo {volume}
  {83}},\ \bibinfo {pages} {587--645} (\bibinfo {year} {2011})}\BibitemShut
  {NoStop}%
\bibitem [{\citenamefont {Anderson}(1995)}]{anderson1995through}%
  \BibitemOpen
  \bibfield  {author} {\bibinfo {author} {\bibfnamefont {Philip~W}\
  \bibnamefont {Anderson}},\ }\bibfield  {title} {\enquote {\bibinfo {title}
  {Through the glass lightly},}\ }\href {\doibase
  https://doi.org/10.1126/science.267.5204.1615.f} {\bibfield  {journal}
  {\bibinfo  {journal} {Science}\ }\textbf {\bibinfo {volume} {267}},\ \bibinfo
  {pages} {1615--1616} (\bibinfo {year} {1995})}\BibitemShut {NoStop}%
\bibitem [{\citenamefont {van Hecke}(2009)}]{van2009jamming}%
  \BibitemOpen
  \bibfield  {author} {\bibinfo {author} {\bibfnamefont {Martin}\ \bibnamefont
  {van Hecke}},\ }\bibfield  {title} {\enquote {\bibinfo {title} {Jamming of
  soft particles: geometry, mechanics, scaling and isostaticity},}\ }\href
  {\doibase https://doi.org/10.1088/0953-8984/22/3/033101} {\bibfield
  {journal} {\bibinfo  {journal} {Journal of Physics: Condensed Matter}\
  }\textbf {\bibinfo {volume} {22}},\ \bibinfo {pages} {033101} (\bibinfo
  {year} {2009})}\BibitemShut {NoStop}%
\bibitem [{\citenamefont {Berthier}\ \emph {et~al.}(2019)\citenamefont
  {Berthier}, \citenamefont {Biroli}, \citenamefont {Charbonneau},
  \citenamefont {Corwin}, \citenamefont {Franz},\ and\ \citenamefont
  {Zamponi}}]{berthier2019gardner}%
  \BibitemOpen
  \bibfield  {author} {\bibinfo {author} {\bibfnamefont {Ludovic}\ \bibnamefont
  {Berthier}}, \bibinfo {author} {\bibfnamefont {Giulio}\ \bibnamefont
  {Biroli}}, \bibinfo {author} {\bibfnamefont {Patrick}\ \bibnamefont
  {Charbonneau}}, \bibinfo {author} {\bibfnamefont {Eric~I}\ \bibnamefont
  {Corwin}}, \bibinfo {author} {\bibfnamefont {Silvio}\ \bibnamefont {Franz}},
  \ and\ \bibinfo {author} {\bibfnamefont {Francesco}\ \bibnamefont
  {Zamponi}},\ }\bibfield  {title} {\enquote {\bibinfo {title} {Gardner physics
  in amorphous solids and beyond},}\ }\href {https://doi.org/10.1063/1.5097175}
  {\bibfield  {journal} {\bibinfo  {journal} {The Journal of chemical physics}\
  }\textbf {\bibinfo {volume} {151}} (\bibinfo {year} {2019})}\BibitemShut
  {NoStop}%
\bibitem [{\citenamefont {Royall}\ \emph {et~al.}(2024)\citenamefont {Royall},
  \citenamefont {Charbonneau}, \citenamefont {Dijkstra}, \citenamefont {Russo},
  \citenamefont {Smallenburg}, \citenamefont {Speck},\ and\ \citenamefont
  {Valeriani}}]{royall2024colloidal}%
  \BibitemOpen
  \bibfield  {author} {\bibinfo {author} {\bibfnamefont {C~Patrick}\
  \bibnamefont {Royall}}, \bibinfo {author} {\bibfnamefont {Patrick}\
  \bibnamefont {Charbonneau}}, \bibinfo {author} {\bibfnamefont {Marjolein}\
  \bibnamefont {Dijkstra}}, \bibinfo {author} {\bibfnamefont {John}\
  \bibnamefont {Russo}}, \bibinfo {author} {\bibfnamefont {Frank}\ \bibnamefont
  {Smallenburg}}, \bibinfo {author} {\bibfnamefont {Thomas}\ \bibnamefont
  {Speck}}, \ and\ \bibinfo {author} {\bibfnamefont {Chantal}\ \bibnamefont
  {Valeriani}},\ }\bibfield  {title} {\enquote {\bibinfo {title} {Colloidal
  hard spheres: Triumphs, challenges, and mysteries},}\ }\href {\doibase
  https://doi.org/10.1103/RevModPhys.96.045003} {\bibfield  {journal} {\bibinfo
   {journal} {Reviews of Modern Physics}\ }\textbf {\bibinfo {volume} {96}},\
  \bibinfo {pages} {045003} (\bibinfo {year} {2024})}\BibitemShut {NoStop}%
\bibitem [{\citenamefont {Carnahan}\ and\ \citenamefont
  {Starling}(1969)}]{carnahan1969equation}%
  \BibitemOpen
  \bibfield  {author} {\bibinfo {author} {\bibfnamefont {Norman~F}\
  \bibnamefont {Carnahan}}\ and\ \bibinfo {author} {\bibfnamefont {Kenneth~E}\
  \bibnamefont {Starling}},\ }\bibfield  {title} {\enquote {\bibinfo {title}
  {Equation of state for nonattracting rigid spheres},}\ }\href {\doibase
  10.1063/1.1672048} {\bibfield  {journal} {\bibinfo  {journal} {Journal of
  chemical physics}\ }\textbf {\bibinfo {volume} {51}},\ \bibinfo {pages}
  {635--636} (\bibinfo {year} {1969})}\BibitemShut {NoStop}%
\bibitem [{\citenamefont {Conway}\ and\ \citenamefont
  {Sloane}(2013)}]{conway2013sphere}%
  \BibitemOpen
  \bibfield  {author} {\bibinfo {author} {\bibfnamefont {John~Horton}\
  \bibnamefont {Conway}}\ and\ \bibinfo {author} {\bibfnamefont {Neil
  James~Alexander}\ \bibnamefont {Sloane}},\ }\href@noop {} {\emph {\bibinfo
  {title} {Sphere packings, lattices and groups}}},\ Vol.\ \bibinfo {volume}
  {290}\ (\bibinfo  {publisher} {Springer Science \& Business Media},\ \bibinfo
  {year} {2013})\BibitemShut {NoStop}%
\bibitem [{\citenamefont {Metropolis}\ \emph {et~al.}(1953)\citenamefont
  {Metropolis}, \citenamefont {Rosenbluth}, \citenamefont {Rosenbluth},
  \citenamefont {Teller},\ and\ \citenamefont
  {Teller}}]{metropolis1953equation}%
  \BibitemOpen
  \bibfield  {author} {\bibinfo {author} {\bibfnamefont {Nicholas}\
  \bibnamefont {Metropolis}}, \bibinfo {author} {\bibfnamefont {Arianna~W}\
  \bibnamefont {Rosenbluth}}, \bibinfo {author} {\bibfnamefont {Marshall~N}\
  \bibnamefont {Rosenbluth}}, \bibinfo {author} {\bibfnamefont {Augusta~H}\
  \bibnamefont {Teller}}, \ and\ \bibinfo {author} {\bibfnamefont {Edward}\
  \bibnamefont {Teller}},\ }\bibfield  {title} {\enquote {\bibinfo {title}
  {Equation of state calculations by fast computing machines},}\ }\href
  {\doibase https://doi.org/10.1063/1.1699114} {\bibfield  {journal} {\bibinfo
  {journal} {The journal of chemical physics}\ }\textbf {\bibinfo {volume}
  {21}},\ \bibinfo {pages} {1087--1092} (\bibinfo {year} {1953})}\BibitemShut
  {NoStop}%
\bibitem [{\citenamefont {Hansen}\ and\ \citenamefont
  {McDonald}(2013)}]{hansen2013theory}%
  \BibitemOpen
  \bibfield  {author} {\bibinfo {author} {\bibfnamefont {Jean-Pierre}\
  \bibnamefont {Hansen}}\ and\ \bibinfo {author} {\bibfnamefont {Ian~Ranald}\
  \bibnamefont {McDonald}},\ }\href@noop {} {\emph {\bibinfo {title} {Theory of
  simple liquids: with applications to soft matter}}}\ (\bibinfo  {publisher}
  {Academic press},\ \bibinfo {year} {2013})\BibitemShut {NoStop}%
\bibitem [{\citenamefont {Bernard}\ \emph {et~al.}(2009)\citenamefont
  {Bernard}, \citenamefont {Krauth},\ and\ \citenamefont
  {Wilson}}]{bernard2009event}%
  \BibitemOpen
  \bibfield  {author} {\bibinfo {author} {\bibfnamefont {Etienne~P}\
  \bibnamefont {Bernard}}, \bibinfo {author} {\bibfnamefont {Werner}\
  \bibnamefont {Krauth}}, \ and\ \bibinfo {author} {\bibfnamefont {David~B}\
  \bibnamefont {Wilson}},\ }\bibfield  {title} {\enquote {\bibinfo {title}
  {Event-chain monte carlo algorithms for hard-sphere systems},}\ }\href
  {\doibase https://doi.org/10.1103/PhysRevE.80.056704} {\bibfield  {journal}
  {\bibinfo  {journal} {Physical Review E—Statistical, Nonlinear, and Soft
  Matter Physics}\ }\textbf {\bibinfo {volume} {80}},\ \bibinfo {pages}
  {056704} (\bibinfo {year} {2009})}\BibitemShut {NoStop}%
\bibitem [{\citenamefont {Engel}\ \emph {et~al.}(2013)\citenamefont {Engel},
  \citenamefont {Anderson}, \citenamefont {Glotzer}, \citenamefont {Isobe},
  \citenamefont {Bernard},\ and\ \citenamefont {Krauth}}]{engel2013hard}%
  \BibitemOpen
  \bibfield  {author} {\bibinfo {author} {\bibfnamefont {Michael}\ \bibnamefont
  {Engel}}, \bibinfo {author} {\bibfnamefont {Joshua~A}\ \bibnamefont
  {Anderson}}, \bibinfo {author} {\bibfnamefont {Sharon~C}\ \bibnamefont
  {Glotzer}}, \bibinfo {author} {\bibfnamefont {Masaharu}\ \bibnamefont
  {Isobe}}, \bibinfo {author} {\bibfnamefont {Etienne~P}\ \bibnamefont
  {Bernard}}, \ and\ \bibinfo {author} {\bibfnamefont {Werner}\ \bibnamefont
  {Krauth}},\ }\bibfield  {title} {\enquote {\bibinfo {title} {Hard-disk
  equation of state: First-order liquid-hexatic transition in two dimensions
  with three simulation methods},}\ }\href {\doibase
  https://doi.org/10.1103/PhysRevE.87.042134} {\bibfield  {journal} {\bibinfo
  {journal} {Physical Review E—Statistical, Nonlinear, and Soft Matter
  Physics}\ }\textbf {\bibinfo {volume} {87}},\ \bibinfo {pages} {042134}
  (\bibinfo {year} {2013})}\BibitemShut {NoStop}%
\bibitem [{\citenamefont {Smallenburg}(2022)}]{smallenburg2022efficient}%
  \BibitemOpen
  \bibfield  {author} {\bibinfo {author} {\bibfnamefont {Frank}\ \bibnamefont
  {Smallenburg}},\ }\bibfield  {title} {\enquote {\bibinfo {title} {Efficient
  event-driven simulations of hard spheres},}\ }\href {\doibase
  https://doi.org/10.1140/epje/s10189-022-00180-8} {\bibfield  {journal}
  {\bibinfo  {journal} {The European Physical Journal E}\ }\textbf {\bibinfo
  {volume} {45}},\ \bibinfo {pages} {22} (\bibinfo {year} {2022})}\BibitemShut
  {NoStop}%
\bibitem [{\citenamefont {Skoge}\ \emph {et~al.}(2006)\citenamefont {Skoge},
  \citenamefont {Donev}, \citenamefont {Stillinger},\ and\ \citenamefont
  {Torquato}}]{skoge2006packing}%
  \BibitemOpen
  \bibfield  {author} {\bibinfo {author} {\bibfnamefont {Monica}\ \bibnamefont
  {Skoge}}, \bibinfo {author} {\bibfnamefont {Aleksandar}\ \bibnamefont
  {Donev}}, \bibinfo {author} {\bibfnamefont {Frank~H}\ \bibnamefont
  {Stillinger}}, \ and\ \bibinfo {author} {\bibfnamefont {Salvatore}\
  \bibnamefont {Torquato}},\ }\bibfield  {title} {\enquote {\bibinfo {title}
  {Packing hyperspheres in high-dimensional euclidean spaces},}\ }\href
  {\doibase https://doi.org/10.1103/PhysRevE.74.041127} {\bibfield  {journal}
  {\bibinfo  {journal} {Physical Review E—Statistical, Nonlinear, and Soft
  Matter Physics}\ }\textbf {\bibinfo {volume} {74}},\ \bibinfo {pages}
  {041127} (\bibinfo {year} {2006})}\BibitemShut {NoStop}%
\bibitem [{git()}]{githubrepo}%
  \BibitemOpen
  \href@noop {} {}\bibinfo {howpublished}
  {\url{https://github.com/mmkafker/hardspheres}}\BibitemShut {NoStop}%
\bibitem [{\citenamefont {Jenks}()}]{sortedcontainers}%
  \BibitemOpen
  \bibfield  {author} {\bibinfo {author} {\bibfnamefont {Grant}\ \bibnamefont
  {Jenks}},\ }\href@noop {} {}\bibinfo {howpublished}
  {\url{https://grantjenks.com/docs/sortedcontainers/}}\BibitemShut {NoStop}%
\bibitem [{\citenamefont {Lubachevsky}\ and\ \citenamefont
  {Stillinger}(1990)}]{lubachevsky1990geometric}%
  \BibitemOpen
  \bibfield  {author} {\bibinfo {author} {\bibfnamefont {Boris~D}\ \bibnamefont
  {Lubachevsky}}\ and\ \bibinfo {author} {\bibfnamefont {Frank~H}\ \bibnamefont
  {Stillinger}},\ }\bibfield  {title} {\enquote {\bibinfo {title} {Geometric
  properties of random disk packings},}\ }\href {\doibase
  https://doi.org/10.1007/BF01025983} {\bibfield  {journal} {\bibinfo
  {journal} {Journal of statistical Physics}\ }\textbf {\bibinfo {volume}
  {60}},\ \bibinfo {pages} {561--583} (\bibinfo {year} {1990})}\BibitemShut
  {NoStop}%
\bibitem [{\citenamefont {Erpenbeck}\ and\ \citenamefont
  {Wood}(1984)}]{erpenbeck1984molecular}%
  \BibitemOpen
  \bibfield  {author} {\bibinfo {author} {\bibfnamefont {Jerome~J}\
  \bibnamefont {Erpenbeck}}\ and\ \bibinfo {author} {\bibfnamefont {William~W}\
  \bibnamefont {Wood}},\ }\bibfield  {title} {\enquote {\bibinfo {title}
  {Molecular dynamics calculations of the hard-sphere equation of state},}\
  }\href {\doibase https://doi.org/10.1007/BF01014387} {\bibfield  {journal}
  {\bibinfo  {journal} {Journal of statistical physics}\ }\textbf {\bibinfo
  {volume} {35}},\ \bibinfo {pages} {321--340} (\bibinfo {year}
  {1984})}\BibitemShut {NoStop}%
\bibitem [{\citenamefont {Hoover}\ and\ \citenamefont
  {Alder}(1967)}]{hoover1967studies}%
  \BibitemOpen
  \bibfield  {author} {\bibinfo {author} {\bibfnamefont {William~G}\
  \bibnamefont {Hoover}}\ and\ \bibinfo {author} {\bibfnamefont {Berni~J}\
  \bibnamefont {Alder}},\ }\bibfield  {title} {\enquote {\bibinfo {title}
  {Studies in molecular dynamics. iv. the pressure, collision rate, and their
  number dependence for hard disks},}\ }\href {\doibase
  https://doi.org/10.1063/1.1840726} {\bibfield  {journal} {\bibinfo  {journal}
  {The Journal of Chemical Physics}\ }\textbf {\bibinfo {volume} {46}},\
  \bibinfo {pages} {686--691} (\bibinfo {year} {1967})}\BibitemShut {NoStop}%
\bibitem [{\citenamefont {Speedy}(1998)}]{speedy1998pressure}%
  \BibitemOpen
  \bibfield  {author} {\bibinfo {author} {\bibfnamefont {RJ}~\bibnamefont
  {Speedy}},\ }\bibfield  {title} {\enquote {\bibinfo {title} {Pressure and
  entropy of hard-sphere crystals},}\ }\href {\doibase
  10.1088/0953-8984/10/20/006} {\bibfield  {journal} {\bibinfo  {journal}
  {Journal of Physics: Condensed Matter}\ }\textbf {\bibinfo {volume} {10}},\
  \bibinfo {pages} {4387} (\bibinfo {year} {1998})}\BibitemShut {NoStop}%
\bibitem [{\citenamefont {Speedy}(1997)}]{speedy1997pressure}%
  \BibitemOpen
  \bibfield  {author} {\bibinfo {author} {\bibfnamefont {Robin~J}\ \bibnamefont
  {Speedy}},\ }\bibfield  {title} {\enquote {\bibinfo {title} {Pressure of the
  metastable hard-sphere fluid},}\ }\href {\doibase 10.1088/0953-8984/9/41/006}
  {\bibfield  {journal} {\bibinfo  {journal} {Journal of Physics: Condensed
  Matter}\ }\textbf {\bibinfo {volume} {9}},\ \bibinfo {pages} {8591} (\bibinfo
  {year} {1997})}\BibitemShut {NoStop}%
\bibitem [{\citenamefont {Speedy}(1994)}]{speedy1994quench}%
  \BibitemOpen
  \bibfield  {author} {\bibinfo {author} {\bibfnamefont {Robin~J}\ \bibnamefont
  {Speedy}},\ }\bibfield  {title} {\enquote {\bibinfo {title} {Quench rate
  independence of the hard sphere glass transition},}\ }\href {\doibase
  https://doi.org/10.1080/00268979400101451} {\bibfield  {journal} {\bibinfo
  {journal} {Molecular Physics}\ }\textbf {\bibinfo {volume} {83}},\ \bibinfo
  {pages} {591--597} (\bibinfo {year} {1994})}\BibitemShut {NoStop}%
\bibitem [{\citenamefont {Torquato}\ \emph {et~al.}(2000)\citenamefont
  {Torquato}, \citenamefont {Truskett},\ and\ \citenamefont
  {Debenedetti}}]{torquatorandomclosepacking}%
  \BibitemOpen
  \bibfield  {author} {\bibinfo {author} {\bibfnamefont {S.}~\bibnamefont
  {Torquato}}, \bibinfo {author} {\bibfnamefont {T.~M.}\ \bibnamefont
  {Truskett}}, \ and\ \bibinfo {author} {\bibfnamefont {P.~G.}\ \bibnamefont
  {Debenedetti}},\ }\bibfield  {title} {\enquote {\bibinfo {title} {Is random
  close packing of spheres well defined?}}\ }\href {\doibase
  10.1103/PhysRevLett.84.2064} {\bibfield  {journal} {\bibinfo  {journal}
  {Phys. Rev. Lett.}\ }\textbf {\bibinfo {volume} {84}},\ \bibinfo {pages}
  {2064--2067} (\bibinfo {year} {2000})}\BibitemShut {NoStop}%
\bibitem [{\citenamefont {Zaccone}(2022)}]{zaccone2022explicit}%
  \BibitemOpen
  \bibfield  {author} {\bibinfo {author} {\bibfnamefont {Alessio}\ \bibnamefont
  {Zaccone}},\ }\bibfield  {title} {\enquote {\bibinfo {title} {Explicit
  analytical solution for random close packing in d= 2 and d= 3},}\ }\href
  {\doibase https://doi.org/10.1103/PhysRevLett.128.028002} {\bibfield
  {journal} {\bibinfo  {journal} {Physical Review Letters}\ }\textbf {\bibinfo
  {volume} {128}},\ \bibinfo {pages} {028002} (\bibinfo {year}
  {2022})}\BibitemShut {NoStop}%
\bibitem [{\citenamefont {Anzivino}\ \emph {et~al.}(2023)\citenamefont
  {Anzivino}, \citenamefont {Casiulis}, \citenamefont {Zhang}, \citenamefont
  {Moussa}, \citenamefont {Martiniani},\ and\ \citenamefont
  {Zaccone}}]{zaccone2023}%
  \BibitemOpen
  \bibfield  {author} {\bibinfo {author} {\bibfnamefont {Carmine}\ \bibnamefont
  {Anzivino}}, \bibinfo {author} {\bibfnamefont {Mathias}\ \bibnamefont
  {Casiulis}}, \bibinfo {author} {\bibfnamefont {Tom}\ \bibnamefont {Zhang}},
  \bibinfo {author} {\bibfnamefont {Amgad~Salah}\ \bibnamefont {Moussa}},
  \bibinfo {author} {\bibfnamefont {Stefano}\ \bibnamefont {Martiniani}}, \
  and\ \bibinfo {author} {\bibfnamefont {Alessio}\ \bibnamefont {Zaccone}},\
  }\bibfield  {title} {\enquote {\bibinfo {title} {Estimating random close
  packing in polydisperse and bidisperse hard spheres via an equilibrium model
  of crowding},}\ }\href {https://doi.org/10.1063/5.0137111} {\bibfield
  {journal} {\bibinfo  {journal} {The Journal of Chemical Physics}\ }\textbf
  {\bibinfo {volume} {158}} (\bibinfo {year} {2023})}\BibitemShut {NoStop}%
\bibitem [{\citenamefont {Zaccone}(2025)}]{zaccone2025complete}%
  \BibitemOpen
  \bibfield  {author} {\bibinfo {author} {\bibfnamefont {Alessio}\ \bibnamefont
  {Zaccone}},\ }\bibfield  {title} {\enquote {\bibinfo {title} {Complete
  mathematical theory of the jamming transition: A perspective},}\ }\href
  {https://doi.org/10.1063/5.0245684} {\bibfield  {journal} {\bibinfo
  {journal} {Journal of Applied Physics}\ }\textbf {\bibinfo {volume} {137}}
  (\bibinfo {year} {2025})}\BibitemShut {NoStop}%
\bibitem [{\citenamefont {Smallenburg}\ \emph {et~al.}(2024)\citenamefont
  {Smallenburg}, \citenamefont {Del~Monte}, \citenamefont {de~Jager},\ and\
  \citenamefont {Filion}}]{smallenburg2024}%
  \BibitemOpen
  \bibfield  {author} {\bibinfo {author} {\bibfnamefont {Frank}\ \bibnamefont
  {Smallenburg}}, \bibinfo {author} {\bibfnamefont {Giovanni}\ \bibnamefont
  {Del~Monte}}, \bibinfo {author} {\bibfnamefont {Marjolein}\ \bibnamefont
  {de~Jager}}, \ and\ \bibinfo {author} {\bibfnamefont {Laura}\ \bibnamefont
  {Filion}},\ }\bibfield  {title} {\enquote {\bibinfo {title} {A simple and
  accurate method to determine fluid–crystal phase boundaries from direct
  coexistence simulations},}\ }\href {\doibase 10.1063/5.0213535} {\bibfield
  {journal} {\bibinfo  {journal} {The Journal of Chemical Physics}\ }\textbf
  {\bibinfo {volume} {160}},\ \bibinfo {pages} {224109} (\bibinfo {year}
  {2024})}\BibitemShut {NoStop}%
\bibitem [{\citenamefont {Donev}\ \emph {et~al.}(2005)\citenamefont {Donev},
  \citenamefont {Torquato},\ and\ \citenamefont {Stillinger}}]{nearlyjammed}%
  \BibitemOpen
  \bibfield  {author} {\bibinfo {author} {\bibfnamefont {Aleksandar}\
  \bibnamefont {Donev}}, \bibinfo {author} {\bibfnamefont {Salvatore}\
  \bibnamefont {Torquato}}, \ and\ \bibinfo {author} {\bibfnamefont {Frank~H.}\
  \bibnamefont {Stillinger}},\ }\bibfield  {title} {\enquote {\bibinfo {title}
  {Pair correlation function characteristics of nearly jammed disordered and
  ordered hard-sphere packings},}\ }\href {\doibase 10.1103/PhysRevE.71.011105}
  {\bibfield  {journal} {\bibinfo  {journal} {Phys. Rev. E}\ }\textbf {\bibinfo
  {volume} {71}},\ \bibinfo {pages} {011105} (\bibinfo {year}
  {2005})}\BibitemShut {NoStop}%
\bibitem [{\citenamefont {Newman}(2018)}]{newman2018networks}%
  \BibitemOpen
  \bibfield  {author} {\bibinfo {author} {\bibfnamefont {Mark}\ \bibnamefont
  {Newman}},\ }\href {\doibase 10.1093/oso/9780198805090.001.0001} {\emph
  {\bibinfo {title} {Networks}}}\ (\bibinfo  {publisher} {Oxford university
  press},\ \bibinfo {year} {2018})\BibitemShut {NoStop}%
\bibitem [{\citenamefont {Wolfram}(2002)}]{Wolfram2002}%
  \BibitemOpen
  \bibfield  {author} {\bibinfo {author} {\bibfnamefont {Stephen}\ \bibnamefont
  {Wolfram}},\ }\href {https://www.wolframscience.com} {\emph {\bibinfo {title}
  {A New Kind of Science}}}\ (\bibinfo  {publisher} {Wolfram Media},\ \bibinfo
  {year} {2002})\BibitemShut {NoStop}%
\bibitem [{\citenamefont {Donev}\ \emph {et~al.}(2004)\citenamefont {Donev},
  \citenamefont {Torquato}, \citenamefont {Stillinger},\ and\ \citenamefont
  {Connelly}}]{donev2004jamming}%
  \BibitemOpen
  \bibfield  {author} {\bibinfo {author} {\bibfnamefont {Aleksandar}\
  \bibnamefont {Donev}}, \bibinfo {author} {\bibfnamefont {Salvatore}\
  \bibnamefont {Torquato}}, \bibinfo {author} {\bibfnamefont {Frank~H}\
  \bibnamefont {Stillinger}}, \ and\ \bibinfo {author} {\bibfnamefont {Robert}\
  \bibnamefont {Connelly}},\ }\bibfield  {title} {\enquote {\bibinfo {title}
  {Jamming in hard sphere and disk packings},}\ }\href {\doibase
  https://doi.org/10.1063/1.1633647} {\bibfield  {journal} {\bibinfo  {journal}
  {Journal of applied physics}\ }\textbf {\bibinfo {volume} {95}},\ \bibinfo
  {pages} {989--999} (\bibinfo {year} {2004})}\BibitemShut {NoStop}%
\bibitem [{\citenamefont {Torquato}\ and\ \citenamefont
  {Stillinger}(2010)}]{torquato2010jammed}%
  \BibitemOpen
  \bibfield  {author} {\bibinfo {author} {\bibfnamefont {Salvatore}\
  \bibnamefont {Torquato}}\ and\ \bibinfo {author} {\bibfnamefont {Frank~H}\
  \bibnamefont {Stillinger}},\ }\bibfield  {title} {\enquote {\bibinfo {title}
  {Jammed hard-particle packings: From kepler to bernal and beyond},}\ }\href
  {\doibase https://doi.org/10.1103/RevModPhys.82.2633} {\bibfield  {journal}
  {\bibinfo  {journal} {Reviews of modern physics}\ }\textbf {\bibinfo {volume}
  {82}},\ \bibinfo {pages} {2633--2672} (\bibinfo {year} {2010})}\BibitemShut
  {NoStop}%
\bibitem [{\citenamefont {Wolfram}(2020)}]{wolfram2020class}%
  \BibitemOpen
  \bibfield  {author} {\bibinfo {author} {\bibfnamefont {Stephen}\ \bibnamefont
  {Wolfram}},\ }\bibfield  {title} {\enquote {\bibinfo {title} {A class of
  models with the potential to represent fundamental physics},}\ }\href@noop {}
  {\bibfield  {journal} {\bibinfo  {journal} {arXiv preprint arXiv:2004.08210}\
  } (\bibinfo {year} {2020})}\BibitemShut {NoStop}%
\bibitem [{\citenamefont {Kronheimer}\ and\ \citenamefont
  {Penrose}(1967)}]{kronheimer1967structure}%
  \BibitemOpen
  \bibfield  {author} {\bibinfo {author} {\bibfnamefont {Erwin~H}\ \bibnamefont
  {Kronheimer}}\ and\ \bibinfo {author} {\bibfnamefont {Roger}\ \bibnamefont
  {Penrose}},\ }\bibfield  {title} {\enquote {\bibinfo {title} {On the
  structure of causal spaces},}\ }in\ \href {\doibase
  https://doi.org/10.1017/S030500410004144X} {\emph {\bibinfo {booktitle}
  {Mathematical Proceedings of the Cambridge Philosophical Society}}},\
  Vol.~\bibinfo {volume} {63}\ (\bibinfo {organization} {Cambridge University
  Press},\ \bibinfo {year} {1967})\ pp.\ \bibinfo {pages}
  {481--501}\BibitemShut {NoStop}%
\bibitem [{\citenamefont {Surya}(2019)}]{surya2019causal}%
  \BibitemOpen
  \bibfield  {author} {\bibinfo {author} {\bibfnamefont {Sumati}\ \bibnamefont
  {Surya}},\ }\bibfield  {title} {\enquote {\bibinfo {title} {The causal set
  approach to quantum gravity},}\ }\href@noop {} {\bibfield  {journal}
  {\bibinfo  {journal} {Living Reviews in Relativity}\ }\textbf {\bibinfo
  {volume} {22}},\ \bibinfo {pages} {1--75} (\bibinfo {year}
  {2019})}\BibitemShut {NoStop}%
\bibitem [{\citenamefont {Arsiwalla}\ and\ \citenamefont
  {Gorard}(2024)}]{arsiwalla2024pregeometric}%
  \BibitemOpen
  \bibfield  {author} {\bibinfo {author} {\bibfnamefont {Xerxes~D}\
  \bibnamefont {Arsiwalla}}\ and\ \bibinfo {author} {\bibfnamefont {Jonathan}\
  \bibnamefont {Gorard}},\ }\bibfield  {title} {\enquote {\bibinfo {title}
  {Pregeometric spaces from wolfram model rewriting systems as homotopy
  types},}\ }\href
  {https://link.springer.com/article/10.1007/s10773-024-05576-0} {\bibfield
  {journal} {\bibinfo  {journal} {International Journal of Theoretical
  Physics}\ }\textbf {\bibinfo {volume} {63:83}} (\bibinfo {year}
  {2024})}\BibitemShut {NoStop}%
\bibitem [{\citenamefont {Arsiwalla}\ \emph {et~al.}(2021)\citenamefont
  {Arsiwalla}, \citenamefont {Gorard},\ and\ \citenamefont
  {Elshatlawy}}]{arsiwalla2021homotopies}%
  \BibitemOpen
  \bibfield  {author} {\bibinfo {author} {\bibfnamefont {Xerxes~D}\
  \bibnamefont {Arsiwalla}}, \bibinfo {author} {\bibfnamefont {Jonathan}\
  \bibnamefont {Gorard}}, \ and\ \bibinfo {author} {\bibfnamefont {Hatem}\
  \bibnamefont {Elshatlawy}},\ }\bibfield  {title} {\enquote {\bibinfo {title}
  {Homotopies in multiway (non-deterministic) rewriting systems as $n$-fold
  categories},}\ }\href {https://arxiv.org/pdf/2105.10822} {\bibfield
  {journal} {\bibinfo  {journal} {arXiv preprint arXiv:2105.10822}\ } (\bibinfo
  {year} {2021})}\BibitemShut {NoStop}%
\bibitem [{\citenamefont {Wolfram}(2023)}]{wolfram2023secondlaw}%
  \BibitemOpen
  \bibfield  {author} {\bibinfo {author} {\bibfnamefont {Stephen}\ \bibnamefont
  {Wolfram}},\ }\href@noop {} {\emph {\bibinfo {title} {The Second Law:
  Resolving the Mystery of the Second Law of Thermodynamics}}}\ (\bibinfo
  {publisher} {Wolfram Media},\ \bibinfo {address} {Champaign, IL},\ \bibinfo
  {year} {2023})\BibitemShut {NoStop}%
\bibitem [{\citenamefont {(2021)}()}]{VertexOutComponentGraph}%
  \BibitemOpen
  \bibfield  {author} {\bibinfo {author} {\bibfnamefont {Wolfram~Research}\
  \bibnamefont {(2021)}},\ }\href@noop {} {\enquote {\bibinfo {title}
  {{VertexOutComponentGraph, Wolfram Language function}},}\ }\bibinfo
  {howpublished}
  {\url{https://reference.wolfram.com/language/ref/VertexOutComponentGraph.html}}\BibitemShut
  {NoStop}%
\bibitem [{\citenamefont {Antal}\ \emph {et~al.}(2008)\citenamefont {Antal},
  \citenamefont {Krapivsky},\ and\ \citenamefont {Redner}}]{collisioncascade}%
  \BibitemOpen
  \bibfield  {author} {\bibinfo {author} {\bibfnamefont {T.}~\bibnamefont
  {Antal}}, \bibinfo {author} {\bibfnamefont {P.~L.}\ \bibnamefont
  {Krapivsky}}, \ and\ \bibinfo {author} {\bibfnamefont {S.}~\bibnamefont
  {Redner}},\ }\bibfield  {title} {\enquote {\bibinfo {title} {Exciting hard
  spheres},}\ }\href {\doibase 10.1103/PhysRevE.78.030301} {\bibfield
  {journal} {\bibinfo  {journal} {Phys. Rev. E}\ }\textbf {\bibinfo {volume}
  {78}},\ \bibinfo {pages} {030301} (\bibinfo {year} {2008})}\BibitemShut
  {NoStop}%
\bibitem [{\citenamefont {Pieprzyk}\ \emph {et~al.}(2024)\citenamefont
  {Pieprzyk}, \citenamefont {Bra\ifmmode~\acute{n}\else \'{n}\fi{}ka},
  \citenamefont {Heyes},\ and\ \citenamefont {Bannerman}}]{transportcoeffs}%
  \BibitemOpen
  \bibfield  {author} {\bibinfo {author} {\bibfnamefont {S\l{}awomir}\
  \bibnamefont {Pieprzyk}}, \bibinfo {author} {\bibfnamefont {Arkadiusz~C.}\
  \bibnamefont {Bra\ifmmode~\acute{n}\else \'{n}\fi{}ka}}, \bibinfo {author}
  {\bibfnamefont {David~M.}\ \bibnamefont {Heyes}}, \ and\ \bibinfo {author}
  {\bibfnamefont {Marcus~N.}\ \bibnamefont {Bannerman}},\ }\bibfield  {title}
  {\enquote {\bibinfo {title} {Revised enskog theory and molecular dynamics
  simulations of the viscosities and thermal conductivity of the hard-sphere
  fluid and crystal},}\ }\href {\doibase 10.1103/PhysRevE.109.054119}
  {\bibfield  {journal} {\bibinfo  {journal} {Phys. Rev. E}\ }\textbf {\bibinfo
  {volume} {109}},\ \bibinfo {pages} {054119} (\bibinfo {year}
  {2024})}\BibitemShut {NoStop}%
\bibitem [{\citenamefont {Garc\'{\i}a~de Soria}\ \emph
  {et~al.}(2024)\citenamefont {Garc\'{\i}a~de Soria}, \citenamefont {Maynar},
  \citenamefont {Gu\'ery-Odelin},\ and\ \citenamefont
  {Trizac}}]{BoltzmannBreather}%
  \BibitemOpen
  \bibfield  {author} {\bibinfo {author} {\bibfnamefont {M.~I.}\ \bibnamefont
  {Garc\'{\i}a~de Soria}}, \bibinfo {author} {\bibfnamefont {P.}~\bibnamefont
  {Maynar}}, \bibinfo {author} {\bibfnamefont {David}\ \bibnamefont
  {Gu\'ery-Odelin}}, \ and\ \bibinfo {author} {\bibfnamefont {Emmanuel}\
  \bibnamefont {Trizac}},\ }\bibfield  {title} {\enquote {\bibinfo {title}
  {Fate of boltzmann's breathers: Stokes hypothesis and anomalous
  thermalization},}\ }\href {\doibase 10.1103/PhysRevLett.132.027101}
  {\bibfield  {journal} {\bibinfo  {journal} {Phys. Rev. Lett.}\ }\textbf
  {\bibinfo {volume} {132}},\ \bibinfo {pages} {027101} (\bibinfo {year}
  {2024})}\BibitemShut {NoStop}%
\bibitem [{\citenamefont {Zwanzig}(2001)}]{zwanzig2001nonequilibrium}%
  \BibitemOpen
  \bibfield  {author} {\bibinfo {author} {\bibfnamefont {Robert}\ \bibnamefont
  {Zwanzig}},\ }\href@noop {} {\emph {\bibinfo {title} {Nonequilibrium
  statistical mechanics}}}\ (\bibinfo  {publisher} {Oxford university press},\
  \bibinfo {year} {2001})\BibitemShut {NoStop}%
\end{thebibliography}%
